\documentclass[useAMS,usenatbib]{mn2e}
\usepackage{graphics}
\usepackage{longtable}
\usepackage{rotating}

\title[testing methods of SPS models]
{Testing three derivative methods of stellar population synthesis models
}

\author[Y. Zhang et. al]
{Yu Zhang,$^{1}$$^,$$^{2}$$^,$$^{3}$$^,$\thanks{E-mail:zhy@ynao.ac.cn}  Zhanwen Han,$^{1}$$^,$$^{3}$ Jinzhong Liu,$^{4}$ Fenghui Zhang,$^{1}$$^,$$^{3}$ Xiaoyu Kang $^{1}$$^,$$^{2}$$^,$$^{3}$\\
$^1$National Astronomical Observatories/Yunnan Observatory, Chinese Academy of Sciences, Kunming, 650011, P.R.China \\
$^2$Graduate University of Chinese Academy of Sciences, Beijing,
100049, P.R.China \\
$^3$Key Laboratory for the Structure and Evolution of Celestial
Objects, Chinese Academy of Sciences\\
$^4$National Astronomical Observatories/Xinjiang Observatory, Chinese Academy of Sciences, Urumqi, 830011, P.R.China \\}

\begin{document}

\pagerange{\pageref{firstpage}--\pageref{lastpage}} \pubyear{2011}

\maketitle

\label{firstpage}

\begin{abstract}
The number of methods used to study the properties of galaxies is
increased, and testing these methods is very important.
Galactic globular clusters (GCs) provide an excellent medium for such
test, because they can be considered as simple stellar populations.
We present ages and metallicities for 40 Galactic GCs as determined from three publicly
available techniques, including colour, Lick-index and
spectrum-fitting methods, based on Bruzual \& Charlot
evolutionary population synthesis (EPS) models. By comparing with the ages obtained from colour-magnitude diagrams (CMDs) and metallicities obtained from spectra of stars, we are able to
estimate the ability of these methods on determination of GCs$^{'}$ parameters, which is absolutely necessary. 
As a result, we find that: (i) for the metallicity, our derived metallicities agree with
those derived from the spectra of stars, Lick-index method is suitable to study metallicity for the stellar population systems in the range of $-1.5\la$[Fe/H]$\la-0.7$ and spectrum-fitting method is suitable to study metallicity for the stellar population systems in the range of $-2.3\la$[Fe/H]$\la-1.5$; (ii) for the age, these three methods have difficulties in age determination, our derived ages are smaller (about 2.0\,Gyr, on average) than the results of CMDs for all these three methods. We use Vazdekis and Maraston models to analyze whether our results are dependent on EPS models, and find that the tendency of these two models is the same as that of Bruzual \& Charlot models. 
Our results are independent of the EPS models.
In addition, our test is based on the old GCs and our conclusions may hold for old stellar population systems.

Besides the age-metallicity degeneracy, we examine the possible effects of other factors (horizontal branch morphology, blue straggler
stars, binary interactions and $\alpha-$enhancement) and give quantitative analysis of the influences of these factors on age determinations (except for $\alpha-$enhancement). For colour and spectrum-fitting methods, the age can be underestimated about $0.0-3.0$\,Gyr, $0.0-2.0$\,Gyr, and $0.0-3.0$\,Gyr due to influences of horizontal branch, blue straggler and binary stars, respectively.
And for Lick-index method, the lower limit of maximal change of age is 6.0\,Gyr, 5.0\,Gyr and 3.0\,Gyr due to influences of horizontal branch, blue straggler
 and binary stars, respectively.

\end{abstract}


\section{Introduction}
Understanding the formation and evolution of galaxies is one of the important questions of cosmology.
With the increased telescope size and improved
instruments, we can obtain multiband photometric and high-resolution spectroscopic data of
extragalactic sources. People can take advantage of these data to study the properties of
galaxies to a maximum extent.

Many techniques have been emerged to study the
properties of galaxies, and evolutionary population synthesis (EPS) is one of these
techniques which has been developed rapidly since it was first
introduced by Tinsley (1968). By EPS method, we can compare the observational data of clusters and galaxies
with EPS models to investigate their properties. So far many EPS models (e.g.
Vazdekis 1999; Schulz et al. 2002; Bruzual \& Charlot 2003,
hereafter BC03; Le Borgne et al. 2004; Zhang et al. 2005a,b; Maraston 2005; Zhang \& Li 2006) emerged and were applied widely
in the study of galaxy properties. Based on these EPS models, people
adopted different methods to study the properties of
stellar population (SP) systems, e.g. using colours to estimate age and metallicity (Dorman, $\rm O^{,}$Connell
\& Rood 2003; James et al. 2006); choosing suitable
Lick/IDS absorbtion-line indices to break the age-metallicity
degeneracy (Worthey 1994; Worthey \& Ottaviani 1997; Trager et
al. 1998); using full spectrum-fitting to match SP system spectra with model spectra 
(Koleva et al. 2008, hereafter K08) and so on.

Now it becomes the most important to test
them and estimate the reliability for these methods, especially
for the fact that a lot of extragalactic
studies are heavily dependent on these methods.
Galaxies are not the ideal test medium for these methods because of
their complex mixture of SPs. On the contrary, Galactic globular clusters
(GCs) are suitable test beds for this purpose. Presumably
formed in a single burst, GCs can be characterized by a single age
(\emph{$\tau$}) and a single metallicity (\emph{Z}), they are considered as the
simple stellar populations (SSPs). The age and metallicity can represent the overall SP
component of a GC.
Furthermore, the ages and metallicities of them can be known from spatially
resolved observations (such as colour-magnitude diagrams,
hereafter CMD and spectroscopy of individual stars), which
provide us a benchmark for this test. Therefore we use the Galactic GCs to
study the reliability of these three methods.

Similar to this work, GCs are also used in other tests, such as testing different models and methods based on
GCs (Mendel, Proctor \& Forbes 2007; Riffel et al.2011; K08).
\begin{itemize}
\item
Mendel et al. (2007) have used Galactic GCs to test three EPS model predictions of age,
metallicity and $\alpha$-element abundances.
Riffel et al. (2011) have employed the near-infrared integrated spectra of
12 Galactic GCs to test near-infrared EPS models.

\item
Meanwhile, K08
has used Galactic GCs to investigate the consistency and reliability of using full
spectrum-fitting on the determination of
SSP equivalent age and metallicity. 

\end{itemize}

Beyond the work of K08, we
use Galactic GCs and BC03 models to investigate the reliability of
three (colour, Lick-index and spectrum-fitting) methods on the estimations of age and metallicity.
Through
comparing these results with those obtained by other techniques (such as
CMDs, spectra of stars), it can provide a valuable examination of these derivative methods of EPS models, and
give a guidance to users of these methods.

The outline of this paper is as follows. We give the
Galactic GC sample data, population synthesis models and stellar spectral library
in Section 2. We analyze the methods used in this work in Section
3. In Section 4, we present the results and discussions of this work. The influences of horizontal branch (HB)
morphology, blue straggler stars (BSs), binary interactions and $\alpha-$enhancement on the age determinations
are given in Section 5. The summary is presented in Section 6.

\section{The Galactic GC sample data, population synthesis models and stellar spectral library}
At first, we describe the Galactic GC sample data and the EPS models in this Section. In order to
discuss the influences of HB morphology, BSs and binary stars on the derived ages,
we introduce the high spatial resolution photometric data and stellar spectral library.

\subsection{The sample data}

In this work, we choose 40 Galactic GCs (which have both photometric
and spectroscopic data) as our sample and use the observed data
from different studies. The spectroscopic data of Galactic
GCs are taken from the spectral library of Schiavon et al. (2005,
hereafter S05). These spectra are obtained from the Cerro Tololo
Inter-American Observatory Blanco 4-m telescope with the
Ritchey-Chretien spectrograph. And these spectra cover the wavelength 3350$-$6430\,{\AA} with
a resolution of about 3.1\,{\AA}.

The photometric data of our Galactic GC sample are taken from the 2010
version of the Harris$'$ catalogue (Harris 1996; http://physwww.physics.mcmaster.ca/$\sim$harris/mwgc.dat).
The selected targets span a wide range of cluster
parameters, including metallicity, HB morphology,
Galactic coordinates and Galactocentric distance. And the main characters of them are listed in Table A1, which are taken from Harris$'$ catalogue.

It is well known that the ages of GCs can be obtained with much higher accuracy by CMDs \citep{de05}, and the metallicities can be obtained by the high resolution spectra of stars \citep{carr09}. In order to give a comparison, we
take the ages of 30 GCs from Salaris \& Weiss (2002), \citet{de05} and \citet{mari09}, which determine ages of GCs from the CMDs, and the ages are mainly taken from \citet{sala02}. The metallicities of all these 40 GCs are taken from \citet{zinn84}, \citet{carr97}, \citet{kraf03} and \citet{carr09}, which determine metallicities of GCs from spectra of stars, and the metallicities are mainly taken from \citet{zinn84}.

\subsection{The stellar population synthesis models}
To test these three methods on the parameter determinations, we adopt the standard BC03 models. The standard BC03 population synthesis models utilize
the Padova 1994 evolutionary tracks, the STELIB library (Le Borgne
et al. 2003) and the Chabrier (2003) initial mass function (IMF) with
stellar mass limits of 0.1 and 100\,M$_\odot$. This set of models
presents the intermediate-resolution spectra ($\sim$\,3\,{\AA}), some spectral feature indices and
colours (based on various systems, including Johnson-Cousins and AB
systems). The SSP models given by BC03 span a wide range of age $\tau$ (0.1\,Myr\,$-$\,20.0\,Gyr)
and metallicity \emph{Z} (0.0001$-$0.05), but
these models provide six metallicities:
[Fe/H]\,=\,$-$2.3,\,$-$1.7,\,$-$0.7,\,$-$0.4,\,$+$0.0, and $+$0.4 (here we adopt
Z$_\odot$\,=\,0.02).

In order to discuss the influence of binary stars on age determinations in Section\,5.3, we also adopt the other two EPS models, one is the models of SSPs (Zhang et al. 2004) and the other is the models of binary stellar populations (BSPs, Zhang et al. 2005a). These models were built on the basis of the $Cambridge$ stellar evolution tracks \citep{eggl71,eggl72,eggl73}, $BaSeL$-2.0 stellar atmosphere models (Lejeune, Cuisinier \& Buser 1997, 1998) and various initial distributions of stars.
The main input parameters of the BSP models are as follows: (i) the initial mass of the primary is chosen from the approximation to the IMF of \citet{mill79} as given by \citet{eggl89}; (ii) the initial secondary-mass distribution, which is assumed to be correlated with the initial primary-mass distribution, satisfies a uniform value, i.e., $n$($q$)=1; (iii) the distribution of orbital separations is taken as constant log\,$a$ for wide binaries and falls off smoothly at close separations; (iv) the eccentricity distribution satisfies a uniform form: $e\,=\,X$.
Other assumptions and more details see Zhang et al. (2004) for SSP models and Zhang et al. (2005a) for BSP models. For those previous models, Zhang et al. adopt the 2.2-version of $BaSeL$ library of \citet{leje97,leje98}, which is a low-resolution library. However, the spectra of S05 library are intermediate resolution and we calculate the Lick indices by degrading to the resolution of Lick system (see Section 4.1), 
we do some modification for these two models, instead of using $BaSeL$ library
we choose the high-resolution ($\sim$\,0.1\,{\AA}, see Section 2.3) B\tiny{LUERED} \normalsize{library} of Bertone et al. (2008) in this work. And other input physics remain unchanged.
\subsection{Other photomethic data and stellar spectral library}
For the purpose of investigating the influences of HB morphology and BSs on age determination, in the first step we should select the HB stars and BSs from CMDs of GCs. CMDs taken from \citet[hereafter P02]{piot02} are used to obtain HB stars and BSs for GCs. 
These data are high spatial resolution photometric data and observed with the Wide-Field Planetary Camera 2 on board of Hubble Space Telescope (HST/WFPC2) camera in the \emph{F}439\emph{W} and \emph{F}555\emph{W} bands. The HST snapshot catalogue contains photometric tables for 74 GCs and the $B$ and $V$ magnitudes
are also obtained from the HST photometry through the
transformations given by P02.
There are 26 clusters in common between the spectral library of S05 and P02 photometric catalogue.

In the second step we should give the corresponding
colours and spectra for each HB star and BS. In order to do this,
we adopt two stellar spectral libraries. One is the 2.2-version of $BaSeL$ library of Lejeune et al. (1997, 1998), which is a low-resolution library. This library provides extensive and homogeneous grids of low-resolution theoretical flux distributions in the range of 91$-$1600\,000\,{\AA} and synthetic \emph{UBVRIJHKLM} colours for a large range of stellar parameters: 50 000\,$\geq$\,\emph{T}$\rm_{eff}$/K\,$\geq$\,2000, 5.50\,$\geq$\,log\emph{g}\,$\geq$\,$-$1.02 and 1.0\,$\geq$\,[Fe/H]\,$\geq$\,$-$5.0. By this library we can obtain the colours for HB stars and BSs. In order to obtain high-resolution spectra for them we adopt the other library,
it is  B\tiny{LUERED} \normalsize{library} of Bertone et al. (2008), which is a
high-resolution (\emph{R}\,$>$\,500 000) library. This library contains 832 theoretical
stellar spectra covering the optical range of
3500$-$7500\,$\rm\AA$ (see Bertone et al. 2008 for
more details). It spans a large range of \emph{T}$\rm_{eff}$ (4000$-$50,000\,K) and log\emph{g} (0.0$-$5.0) at six  metallicity values ([Fe/H] $= -3.0, -2.0, -1.0, -0.3, +0.0$ and $+0.3$).

\section{methods}
Having described the sample data and EPS models, we now begin to
introduce the three methods used in this paper, including colour, Lick-index and spectrum-fitting methods.

\subsection{Colour method}

As shown by Yi et al.
(2004), the $U-B$ and $B-V$ colours can be well used to
derive the ages and metallicities for GCs. 
Anders et al. (2004) also
analyzed the reliability and limitation of the combination of various
passbands and they found that the $U$ and $B$ bands are important
for the age and metallicity determinations, and the \emph{V} band and
near-infrared data can provide additional constraints. They listed the
preferable passband combinations in their Tables 1 and 2, and
 the \emph{UBVI} are the preferable combinations for the SP
systems with age larger than a few Gyr from their Table 2. So we select
three colours ($U-B$, $B-V$ and $V-I$) from Harris'
catalogue to fit with models to obtain the ages and
metallicities of GCs. But for the other SP systems, the selection of colours may be different, such as \citet{mara01} have found that the ages of young star clusters can be well determined based on $B-V$ and $V-K$ colours, and wider photometric data can be used well to constrain the parameters for distant SP systems \citep{barr11}. In this procedure, we set the ages and metallicities of SPs free and select the best-fitting model by minimizing the
$\chi^2$ following the formula:

\begin{equation}
\;\;\;\;\;\;\;\;\;\;\;\;\;\chi^2_{i}=
\frac{1}{N\rm_{dof}}\sum\limits_{n=1}^{3}\left({{C\rm_{M\emph{\tiny,i,n}}-\emph{C}\rm_{O\emph{\tiny,n}}}\over{\sigma\rm_{O\emph{\tiny,n}}}}\right)^2,
\end{equation}
where the \emph{C}$\rm_{M}$$_{,i,n}$ is the \emph{n}-th colour corresponding to the
\emph{i}-th SSP model. The \emph{C}$\rm_{O}$$_{,n}$ and $\sigma$$\rm_{O}$$_{,n}$ are the \emph{n}-th  colour
and error of the Galactic GCs.
\emph{N}$\rm_{dof}$ is the degree of freedom. Error
$\sigma$$\rm_{O}$$_{,n}$ is not given in the Harris$'$ catalogue, so we assume
that the error of $U-B$, $B-V$ and $V-I$ is 0.04 mag.

Errors on the age and metallicity estimation are determined from the
$\chi^2$ contours. Because we use three colours in this work, the
\emph{N}$\rm_{dof}$ is 2. And for \emph{N}$\rm_{dof}$\,$=$\,2,
the 1$\sigma$, 2$\sigma$ and 3$\sigma$ errors correspond to the
formal confidence intervals of 68.7\%
($\Delta\chi^2$\,=\,2.3), 95.4\% ($\Delta\chi^2$\,=\,6.17) and
99.73\% ($\Delta\chi^2$\,=\,11.8; Avni 1976), respectively.
\begin{figure}
\begin{flushleft}
\includegraphics[bb=580 700 110 35,height=8.0cm,width=6.cm,clip,angle=90]{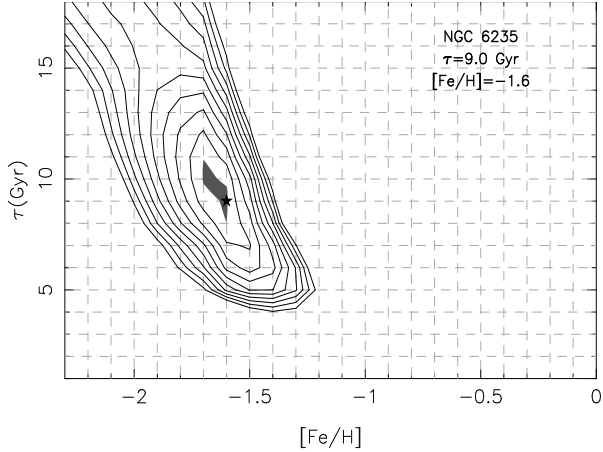}
\caption{Contour map of $\chi^2$ as a function of age $\tau$ and
metallicity [Fe/H] for an example of NGC 6235. From the center to the edge the value of
$\chi^2$ increases from 1.1 to 3.3 in interval of 0.22. The black star stands for the minimum $\chi^2$ and corresponds to the obtaining $\tau$ and [Fe/H]. We list the parameters of NGC 6235 on the upper right corner.
The gray shadow represent 1$\sigma$ error contour, and the errors are obtained based on this gray shadow. The
dashed lines mark the BC03 model grids, we do interpolation for the metallicity (see Section 4.1).}\label{entropy-rlof}
\end{flushleft}
\end{figure}
\begin{figure}
\begin{flushleft}
\includegraphics[bb=80 35 470 785,height=8.5cm,width=4cm,clip,angle=270]{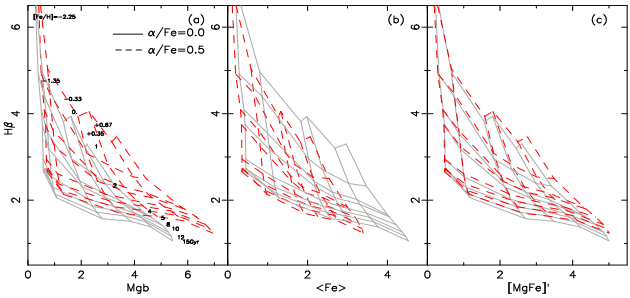}
\caption{H$\beta$ versus three indices (Mg\,$b$, $<$Fe$>$ and [MgFe]$^{'}$) for the models of Thomas, Maraston \& Johansson (2011a), the ages are from $1$ to $15$ Gyr and [Fe/H] are from $-2.3$ to $0.67$ that are labeled in the figure. The gray solid and red dashed lines stand for [$\alpha$/Fe]$=0.0$ and 0.5, respectively.}\label{entropy-rlof}
\end{flushleft}
\end{figure}

Fig.\,1 shows an example of the method of using contour map of $\chi^2$ to obtain age and metallicity for NGC\,6235.
The star stands for the $\chi^2_{min}$ and corresponds to the obtaining $\tau$ and [Fe/H].
The shaded area maps 1$\sigma$ confidence region, and the $\Delta$$\tau$
and $\Delta$[Fe/H] are obtained based on this shaded area.

\subsection{Lick-index method}

Lick/IDS indices have been used to break the age-metallicity
degeneracy (Trager et al. 2000; Terlevich \& Forbes 2002; Gallazzi
et al. 2005) of SP systems. In this paper, we select the Balmer (H$\beta$, H$\delta_{A}$, H$\delta_{F}$,
H$\gamma_{A}$ and H$\gamma_{F}$), Mg\,\emph{b}, Fe5270 and Fe5335
indices to study the parameters of GCs (Balmer indices are age-sensitive indices,
Mg\,\emph{b}, Fe5270 and Fe5335 are metal-sensitive indices). The H$\delta$ and H$\gamma$ are high order 
Balmer indices which measured with a narrower ($\sim$\,20\,${\rm\AA}$; marked by subscript F) and a wider ($\sim$\,40\,${\rm\AA}$; marked by subscript A) central bandpass \citep{wort97}.
Among these five Balmer indices, H$\beta$ and H$\gamma_{F}$ are little sensitive to $\alpha/$Fe and the other three indices are sensitive to $\alpha/$Fe. And in Section 5.4, we will discuss the influence of $\alpha$-enhancement on age determination for Lick-index method.
For
the three metal-sensitive indices, we adopt the following
definition:
[MgFe]$^{'}$\,=\,[Mg\,\emph{b}(0.72\,$\times\,$Fe5270\,+\,0.28\,$\times$\,Fe5335)]$^{1/2}$,
which defined by Thomas, Maraston \& Bender (2003). The [MgFe]$^{'}$ has been found to be good tracer of metallicity. 
Adopting the models of Thomas, Maraston \& Johansson (2011a) we present this character in Fig.\,2, the grids of H$\beta$ versus Mg\,$b$, $<$Fe$>$ ($<$Fe$>=$1/2(Fe5237$+$Fe5335)) and [MgFe]$^{'}$ with two types of $\alpha$/Fe ([$\alpha$/Fe]$=0.0$ and $0.5$), the gray solid and red dashed lines stand for [$\alpha$/Fe]$=0.0$ and $0.5$, respectively. From them we can see that the influence of $\alpha$/Fe ratio is obvious for H$\beta$ versus Mg\,$b$ and $<$Fe$>$ grids, and is small for H$\beta$ versus [MgFe]$^{'}$ grid. The Mg\,$b$ increases and $<$Fe$>$ decreases with increasing $\alpha$/Fe, and the combination index [MgFe]$'$ counteract the effect of $\alpha$/Fe ratio on Mg$b$, Fe5237 and Fe5335. This certify that the [MgFe]$'$ is independent of $\alpha$/Fe and can be considered as good tracer of metallicity \citep{thom03}.

\begin{figure*}
\includegraphics[bb=20 20 600 780,height=16cm,width=9.5cm,clip,angle=270]{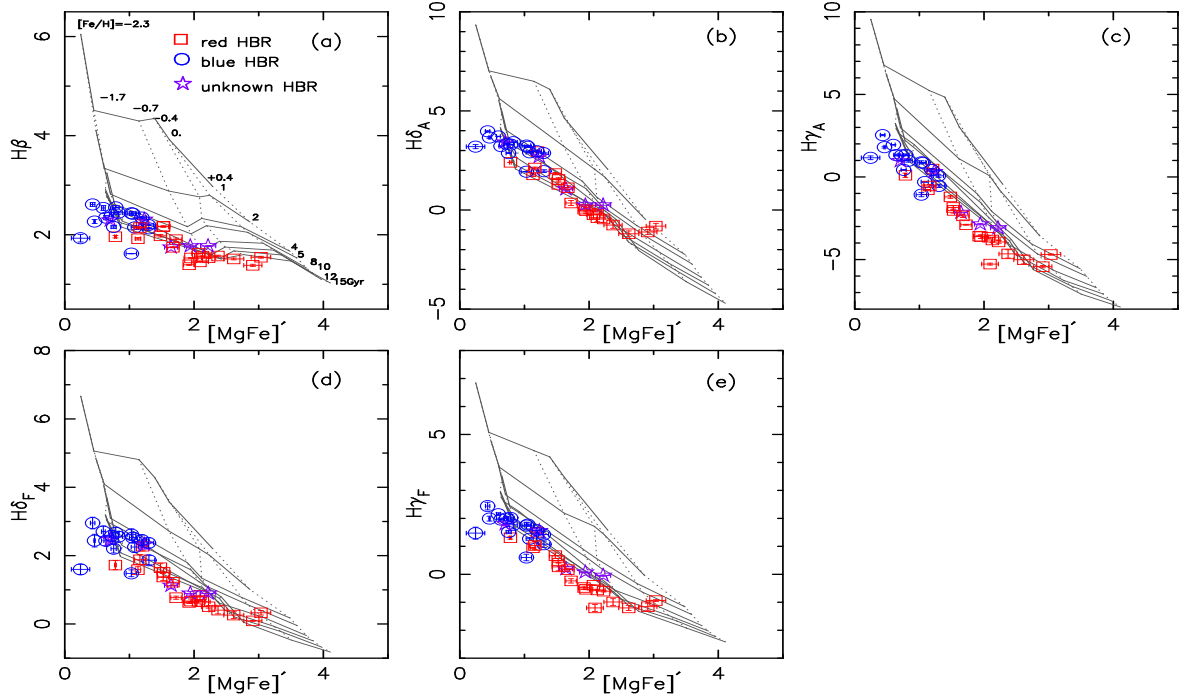}
\caption{The distribution of GCs on the Balmer indices and [MgFe]$^{'}$ planes. Model grides from 
BC03 are shown: age is constant for solid line
(top to bottom, 1, 2, 4, 5, 8, 10, 12 and 15 Gyr), and metallicity is constant for dashed line
(left to right:
[Fe/H]$=-2.3, -1.7, -0.7, -0.4$, $0.0$ and $+0.4$).
Line strengths of Galactic GCs are plotted on SSP grides. The red open squares, blue open circles and purple open pentacles stand for GCs with red, blue and unknown HBR, respectively. 
In the top left plane (H$\beta$
versus [MgFe]$^{'}$ plane), we give these values of age and
metallicity.}\label{entropy-rlof}

\end{figure*}
In Fig.\,3, we show the distribution of GCs on
five Balmer (H$\beta$, H$\delta_{A}$, H$\delta_{F}$, H$\gamma_{A}$ and H$\gamma_{F}$)
indices versus [MgFe]$^{'}$ planes. We can see that the majority of GCs lie beyond the grid covered by H$\beta$ (this problem does not involve all Balmer indices, Thomas et al. 2011a), which would be inconsistent with
the typical universe age 13.6\,Gyr. This is a general problem (see Fig.\,2 of
Mendel et al. 2007) for all studies of using H$\beta$ index to
study the GCs based on existing models. This problem is also clearly demonstrated by the $'$zero-point$'$
offset (some GCs lie beyond the grid of H$\beta$ and most GC ages are greater than 14\,Gyr).
Although these problems have been pointed by some groups
(Vazdekis et al. 2001; Schiavon et al. 2002; Cenarro et al.
2008), but few works have been done on studying the specific sources of this
problem or quantified their effects on the derived ages and metallicities. 
Similar to the technique described in Puzia et al. (2005), the ages and metallicities for individual GCs are computed as the weighted mean of the parameters derived from five Balmer indices versus [MgFe]$^{'}$ grids.

Due to the influence of hot blue HB stars, the isochrones for old ($\tau$\,$>$\,8.0\,Gyr)
stellar populations with metallicity below [Fe/H]$\,=-0.7$ tend
to overlap (Maraston \& Thomas 2000; Mendel et al. 2007), this nature is reflected by H$\beta$ in panel (a) of Fig.\,3.
This brings an ambiguity in determining ages and
metallicities of GCs and unnaturally extends the age
distributions for old ages for Lick-index method.
Note that, for the GCs lying beyond
the Lick indices grids, we assume the values of GCs are 15.0\,Gyr.

\subsection{spectrum-fitting method}
With the increase of spectroscopic data, people use spectra to study the properties of SP
systems. So far there are many kinds of full spectrum-fitting
techniques used in such study (e.g. Cid Fernandes et al.
2005; Mathis, Charlot \& Brinchmann 2006; Tojeiro et al. 2007; K08).

We employ a modified version of the \tiny{STARLIGHT} \normalsize{code} by Cid Fernandes et al.
(2005) to analyze the observed spectra of Galactic GCs. 
The \tiny{STARLIGHT} \normalsize{}code is originally used to study the properties
of galaxy, and is achieved by fitting the observed spectrum \emph{F$\rm_{O}$} with a model spectrum
\emph{F$\rm_{M}$} that mixed by \emph{N$_{\star}$} SSPs with
different ages and metallicities from the BC03 models. The code is carried out with a simulated
annealing plus Metropolis scheme (Cid Fernandes et al. 2001), which
searches for the minimum

\begin{equation}
\;\;\;\;\;\;\;\;\;\;\;\;\;\chi^2= \sum\limits_{\lambda}\,
[(F\rm_{O}-\emph{F}\rm_{M})\omega_{\lambda}]^2,
\end{equation}
where $\omega$$^{-1}_{\lambda}$ is the error in
\emph{F$\rm_{O}$}. The
line-of-sight stellar motions are modeled by a Gaussian distribution
centered at velocity \emph{v}$_{\star}$ and with dispersion
\emph{$\sigma$}$_{\star}$. The output includes the distributions of
stellar age, metallicity, extinction, velocity dispersion and
stellar mass.

\section{Results and Discussions}
We use colour, Lick-index and spectrum-fitting methods based on BC03 model to
obtain the ages and metallicities for Galactic GCs, and compare them with those 
determined from spatially resolved observations (such as CMDs, spectra of stars). 
We also compare the ages and metallicities derived from other two models 
(Vazdekis and Maraston models) to test whether our results are dependent on EPS models.
Before displaying the results of this work, we should introduce some pretreatments of model,
sample data and methods. The details are given as follows.

\subsection{Pretreatments}

\begin{itemize}
\item
For the EPS model, as said above, the BC03 model provides six metallicities, we
linearly interpolate colours, Lick indices and spectra of SSPs that span [Fe/H] from $-$2.3 to $+$0.4 in increments of 0.1 dex.

\item
In order to discuss the influence of HB morphology on age determinations in Section 5.1,
we divide the GCs into three groups according to the value of HB ratio (HBR
\footnote{ HBR\,$=(B-R)/(B+R+V)$, B is the number of blue HB stars, R is the number of red HB stars
and V is the number of RR Lyrae variable.}) from the Harris$'$ catalogue. One is the blue HBR type (HBR\,$>$0),
the second is red HBR type ( HBR\,$<$0) and the third is unknown
HBR (HBR is not given).

\item For the colour method, we do extinction corrections for observed magnitudes of GCs by adopting the extinction curves of \citet{schl98} and
the $E(B-V)$ of Harris$'$ catalogue (the 8th column of Table A1).

\item For the Lick-index method, in this work
we calculate the Lick indices from the spectra of Galactic GCs by degrading to
wave-dependent resolution of the Lick system (see Section 4.4 of BC03), without adopting the results by using
fitting functions. 

\begin{figure*}
\includegraphics[bb=20 20 580 730,height=16.0cm,width=9.5cm,clip,angle=270]{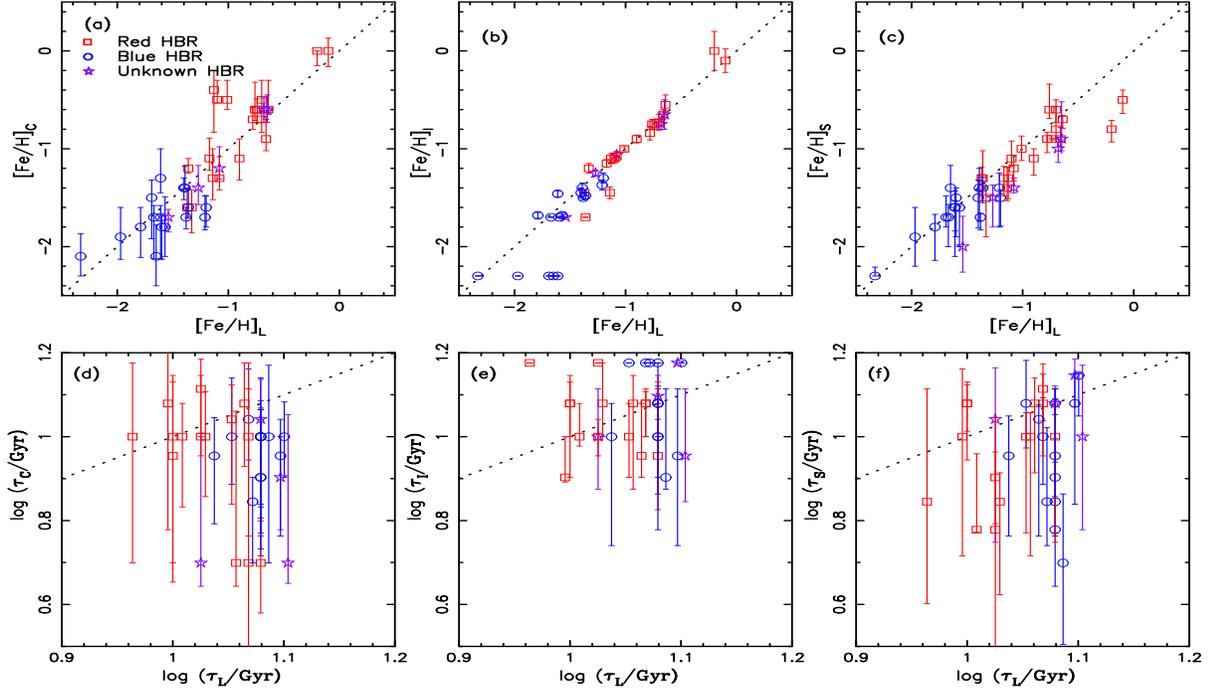}
\caption{Top panels: the comparisons between the metallicities obtained by three
methods and the spectra of stars for Galactic GCs. Symbols have the same meanings in Fig.3. Bottom panels are the comparisons
between the ages obtained by three
methods and the CMDs. Left, middle and right panels are comparison results for methods of colour, Lick-index and spectrum-fitting, respectively.
}\label{entropy-rlof}

\end{figure*}

\item For the spectrum-fitting method, we adopt a modified version of the \tiny{STARLIGHT}\normalsize{} code.
Because it is generally accepted
that the GC can be represented by a SSP, only one
component in the base. We fit the observed GC spectrum
\emph{F$\rm_{O}$} with each of the SSP spectrum \emph{F$_{i}$}, do
not fit the observed spectrum with a linear combination of
\emph{N$_{\star}$} SSPs. In this study, following \citet{cid05}, we find the best fitting SSP, which matches a given observed spectrum of GC, 
through a standard $\chi^2$ minimization procedure:

\begin{equation}
\;\;\;\;\;\;\;\;\;\;\;\;\;\chi^2_{i}= \sum\limits_{\lambda}\,
[(F\rm_{O}-\emph{F$_{i}$})\omega_{\lambda}]^2,
\end{equation}
instead of equation (2), where \emph{F$_{i}$} corresponds to the spectrum of i-th SSP model.
And in this work we construct a base of $18$\,ages\,$\times\,24$\,metallicities\,$=432$\,SSP model spectra. The use of \tiny{STARLIGHT}\normalsize{} to study the integrated spectra of clusters
has also been extensively discussed by Cid Fernandes \&
Gonz$\acute{a}$lez Delgado (2010).

In the spectrum-fitting process, we use the spectra in the 3700$-$5700\,${\rm\AA}$
range because the red and blue parts of the
observations have poor quality. Errors on the age and metallicity
estimation are also determined by the $\chi^2$ contours.

\end{itemize}

\subsection{Comparison of parameters with other studies }

Using the methods and pretreatments outlined in Sections\,3 and 4.1, we obtain ages, metallicities 
and corresponding errors for GCs.
In order to distinguish the ages and metallicities derived from different methods, we define some parameters in this paper.
$\tau\rm_{C}$ and [Fe/H]$\rm_{C}$ stand for the age and metallicity determined from colour method,
$\tau\rm_{I}$ and [Fe/H]$\rm_{I}$ represent those obtained from Lick-index method,
and $\tau\rm_{S}$ and [Fe/H]$\rm_{S}$ stand for those determined from spectrum-fitting method.
The parameters of literatures are represented by $\tau\rm_{L}$ and [Fe/H]$\rm_{L}$.
In Table A2 we list the GC IDs in the first column. In the 2nd to 7th columns, $\tau\rm_{C}$, [Fe/H]$\rm_{C}$,
$\tau\rm_{I}$, [Fe/H]$\rm_{I}$, $\tau\rm_{S}$ and [Fe/H]$\rm_{S}$ are given with errors, respectively.
And in 8th and 9th columns we list the $\tau\rm_{L}$ and [Fe/H]$\rm_{L}$.

In the top panels of Fig.\,4, we give the comparison between our GC metallicities and those given in the literatures that obtain metallicities from the spectra of stars. And 
(a), (b) and (c) panels are the comparison results for colour, Lick-index and spectrum-fitting methods, respectively. The vertical bars are the errors in metallicities derived from three methods. From them
we see that the metallicities obtained by three methods are in agreement with literature value for the whole sample. Meanwhile we also find that all GCs with blue HBR have lower metallicity than those with red HBR for these three methods, which agrees
with the study of Lee, Yoon \& Lee (2000). Our specific analysis are given as follows.\\
(a) For colour method, [Fe/H]$\rm_{C}$ agrees with [Fe/H]$\rm_{L}$ for the whole sample, but there exists discreteness. We can see that [Fe/H]$\rm_{C}$ is a bit smaller than [Fe/H]$\rm_{L}$ in the range of [Fe/H]$\rm_{L}$$\la-$1.0 and [Fe/H]$\rm_{C}$ agrees with [Fe/H]$\rm_{L}$ for the range of [Fe/H]$\rm_{L}$$>-$1.0. But for the value of [Fe/H]$\rm_{L}$ is about $-$1.1, there exist three GCs (NGC 6171, 6342 and 6652) having lager [Fe/H]$\rm_{C}$ than [Fe/H]$\rm_{L}$. From the [Fe/H]$\rm_{C}$ and $\tau\rm_{C}$ of these GCs listed in Table A2, we see that $\tau\rm_{C}$ is smaller than $\tau\rm_{L}$, so these GCs can be affected by age-metallicity degeneracy.\\
(b) For the Lick-index method, the [Fe/H]$\rm_{I}$ is perfectly in agreement with [Fe/H]$\rm_{L}$ in the range of $-1.5\la$\,[Fe/H]\,$\la-0.7$. But there have difficulties in metallicity determination for the range of [Fe/H]\,$\la-1.5$, because it tends to overlap for Balmer and [MgFe]$^{'}$ indices at low-metallicity range for BC03 model. And we can see this phenomenon clearly in the panel (a) of Fig.\,3. This indicates that Lick-index method is suitable to study metallicity for SP systems in the range of $-1.5\la$\,[Fe/H]\,$\la-0.7$ and has some difficulties in studying SP systems for the range of [Fe/H]\,$\la-1.5$.\\
(c) For the spectrum-fitting method, we find that [Fe/H]$\rm_{S}$ matches [Fe/H]$\rm_{L}$ in the range of $-2.3\la$\,[Fe/H]\,$\la-1.5$, and [Fe/H]$\rm_{S}$ is smaller than [Fe/H]$\rm_{L}$ for the range of [Fe/H]\,$\ga-1.5$. All these results imply that the spectrum-fitting method may be suitable to study metallicity for SP systems in the range of $-2.3\la$\,[Fe/H]\,$\la-1.5$ and [Fe/H]$\rm_{S}$ may be smaller than [Fe/H]$\rm_{L}$ for the range of [Fe/H]\,$\ga-1.5$.

On the whole, our metallicities obtained from three methods match those determined from spectra of stars in the entire metallicity range spanned by the GCs. The Lick-index method is suitable to study the metallicity in the range of $-1.5\la$\,[Fe/H]\,$\la-0.7$ and spectrum-fitting method is suitable to study the metallicity in the range of $-2.3\la$\,[Fe/H]\,$\la-1.5$.

In the bottom panels of Fig.\,4, we compare our derived GC ages of three methods with the values of literature that determine ages from CMDs. And (d), (e) and (f) panels are corresponding to colour, Lick-index and spectrum-fitting methods, respectively. From them we can see there exists large discrepancy in age determinations, and the errors are also large for these three methods. Our specific analysis are given as follows.\\
(d) For the colour method, most of the GCs with blue and unknown HBR have relative lower $\tau\rm_{C}$ than $\tau\rm_{L}$, but there has no obvious tendency for those with red HBR. The age of GCs with blue HBR can be influenced by HB stars for the existence of HB stars making GCs look younger. Five GCs are found to possess extreme low $\tau\rm_{C}$, including two with unknown HBR (NGC 6388 and 6441) and three with red HBR (NGC 6171, 6342, and 6652). Among these GCs, three have relative lager [Fe/H]$\rm_{C}$ than [Fe/H]$\rm_{L}$, which can be affected by the age-metallicity degeneracy (see Fig.\,4a). Meanwhile from the 4th and 5th columns of Table A1, we know that the other two GCs are close to the Galactic bulge, which are badly contaminated by the field stars or affected by differential reddening, so we can not get reliable parameters for them.\\
(e) For the Lick-index method, we can see there exist some GCs with extreme large $\tau\rm_{I}$ (about 15.0\,Gyr), and except these GCs most of GCs have smaller $\tau\rm_{I}$ which are influenced by HB stars, especially for those with blue HBR. About eight GCs have large $\tau\rm_{I}$ of 15.0\,Gyr which lie beyond the grids, including two with red HBR (NGC 1851 and 6637), one with unknown HBR (NGC 5946) and five with blue HBR (NGC 1904, 2298, 3201, 6254 and 7078). Similar to above, Lick-index method has some difficulties in parameter determinations for the range of [Fe/H]\,$\la-1.5$, and most of these eight GCs are metal-poor ([Fe/H]$\rm_{I}$$\la-$1.5), so we obtain extreme large $\tau\rm_{I}$ for them.\\
(f) For the spectrum-fitting method, most GCs with blue HBR have relative smaller $\tau\rm_{S}$ than $\tau\rm_{L}$, but there has no obvious tendency for those with red and unknown HBR. From panel (c), we know that [Fe/H]$\rm_{S}$ is consistent with [Fe/H]$\rm_{L}$ in the range of [Fe/H]\,$\la-1.5$ and most of GCs with blue HBR are metal-poor, so the $\tau\rm_{S}$ of GCs with blue HBR are affected by HB stars rather than age-metallicity degeneracy. For the GCs with red and unknown HBR, one part has large $\tau\rm_{S}$ which can be influenced by age-metallicity degeneracy, and the other part has small $\tau\rm_{S}$ which may be influenced by HB stars and BSs.
\begin{table*}
\begin{center}
\caption[]{Comparison of main model ingredients in each EPS model. 
The first, second and third rows list the stellar evolution track, 
stellar spectral library and IMF of the EPS models.
The 4th and 5th rows show the metallicity ($Z$) and age range that these models cover,
and the number of metallicities is also given in the parenthesis of the 4th row.  }
\tabcolsep=0.15in

\label{Tab:frac}
\renewcommand{\arraystretch}{1.5}
\begin{tabular}{|l|l|l|l|r|r|r|r|r|r|c|c|c|c|c|c|c|c|c|c|}
\hline\hline\noalign{\smallskip}
Models &BC03   &Vazdekis &Maraston \\
\hline\noalign{\smallskip}
Stellar evolution track   &Padova 1994  &Padova 2000  &Cassisi+Geneva \\
Stellar spectral library   &STELIB &MILES  &MILES\\
IMF                       &Chabrier &Salpeter  &Chabrier\\
$Z$(number)               &$0.0001-0.0500(6)$&$0.0001-0.0300(7)$&$0.0001-0.0400(5)$\\
Age(Gyr)                  &$0.0001-20.0000$  &$0.0630-17.1800$  &$0.0060-15.0000$\\

\hline
\end{tabular}
\end{center}
\end{table*}
Note that, relevant to this test is that CMD-derived ages also can carry their own problems for this test, because the CMD-derived ages depend on the adopted tracks and on whether element ratios be taken into account, which has also been discussed in \citet{mara11}.

On the whole, all of these three methods have difficulties in age determination 
and we can not directly say which method is better on age determination. 
Except the influences of CMD-derived ages and age-metallicity 
degeneracy in the entire age range, 
there exist some uncertainties in age
determinations, some factors (e.g. HB stars, BSs, binary stars and $\alpha$-enhancement) 
can affect the age determinations and we will discuss them in the next Section.

\begin{figure*}
\includegraphics[bb=20 15 580 770,height=16.0cm,width=9.5cm,clip,angle=270]{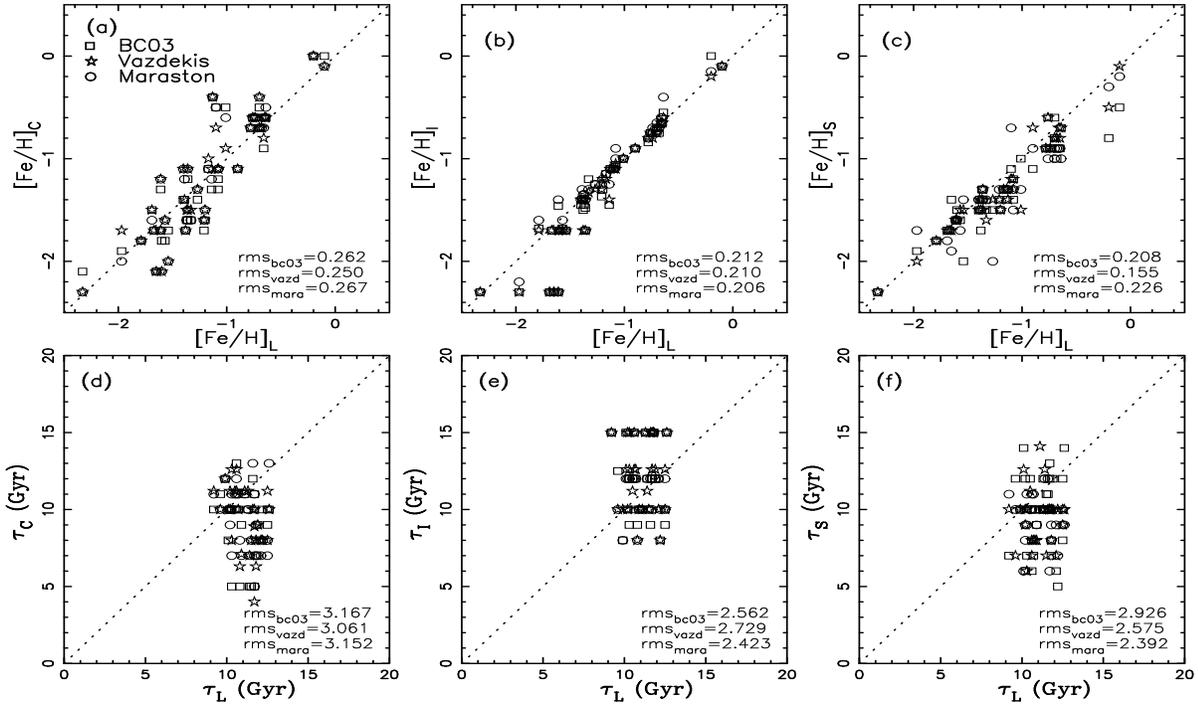}

\caption{Comparisons between the results from three EPS models and the literatures. Top and bottom panels are the comparisons for metallicity and age, respectively. The open squares, pentacles and circles stand for the results of BC03, Vazdekis and Maraston EPS models, respectively. Left, middle and right panels are comparison results for methods of colour, Lick-index and spectrum-fitting, respectively. The dispersion (rms) of each model is also listed in the lower right corner of each panel.
}\label{entropy-rlof}

\end{figure*}

\subsection{Consistency checks}
We use Vazdekis\footnote{http://miles.iac.es/} \citep[the updated v9.1 version,][]{vazd10,falc11} and Maraston \citep{mara11} models to investigate whether our results are dependent on EPS models.
The Vazdekis models use the Padova 2000 theoretical isochrones \citep{gira00}, the MILES library \citep{sanc06} and four types of IMF with stellar mass limits (Vazdekis et al. 2003) of 0.1 and 100 M$_{\odot}$. This version covers the wavelength range of 3540$-$7410\,${\rm\AA}$ with a spectral resolution (FWHM) of $\sim$2.5\,${\rm\AA}$ \citep{falc11}, seven metallicities $0.0001\leqslant Z \leqslant0.03$ and 50 ages across the range of 0.0630$-$17.1800\,Gyr. In this work, we select the unimodal Salpeter IMF \citep{salp55} to study the parameters of GCs for spectrum-fitting and Lick-index methods. For the colour method, because the MILES library has relatively limited wavelength coverage and does not extend into I-band, we select the colours from previous models of Vazdekis \citep{vazd99} to study the parameters of GCs. This does not influence our results (private discussion).

The Maraston models assume a stellar evolution prescription, which consists of the isochrones and stellar tracks by Cassisi, Castellani \& Castellani (1997) for ages larger than $\sim$30.0Myr and by Ganeva \citep{scha92} for younger populations. Sets of models have been also computed with Padova stellar evolutionary models \citep{gira00}. Four different libraries of flux-calibrated empirical stellar spectra and three types of IMF have been considered. We select the MILES library and the \citet{chab03} IMF to study the parameters of GCs. These updated models cover the wavelength range of 3500$-$7430\,${\rm\AA}$ with a spectral resolution (FWHM) of $\sim$2.54\,${\rm\AA}$ \citep{beif11} and five metallicities 
$0.0001\leqslant Z \leqslant0.04$ with different ages across the range of 0.0060$-$15.0000\,Gyr (Maraston \& Str\"{o}mb\"{a}ck 2011, and references therein). Just as said above, we use the updated models of \citet{mara11} based on MILES library for spectrum-fitting and Lick-index methods. And for the colour method we select the models of \citet{mara05}, which based on the $BaSeL$ library, to study the parameters of GCs. In Table 1 we list the comparison of main model ingredients in each EPS model. The first, second and third rows list the stellar evolution track, 
stellar spectral library and IMF of the EPS models that we used in this study.
The 4th and 5th rows show the metallicity ($Z$) and age range that these models cover,
and the number of metallicities is also given in the parenthesis of the 4th row.

We use these two EPS models to obtain the parameters of GCs based on three methods described above and compare the results of these two models with that of the literatures. In Fig.\,5, we show the comparisons between the parameters of GCs obtained from tree methods based on three EPS models (BC03, Vazdekis and Maraston) and the literatures, in which the open squares, pentacles and circles stand for the results of BC03, Vazdekis and Maraston EPS models, respectively. The dispersion (rms) is also listed in the lower right corner of each panel, rms$\rm_{bc03}$, rms$\rm_{vazd}$ and rms$\rm_{mara}$ stand for the rms of BC03, Vazdekis and Maraston EPS models, respectively. We find that the metallicities obtained from three methods of these two EPS models (Vazdekis and Maraston) have an agreement with those of the literatures in the entire metallicity range. From the top panels (a, b and c panels), we find that the comparisons between metallicities obtained by three methods and the spectra of stars for these two models are similar to that of BC03 models. For Lick-index method in panel (b), we see that the rms$\rm_{vazd}$ and rms$\rm_{mara}$ are smaller than rms$\rm_{bc03}$, this may be due to the adoption of MILES for these two models and STELIB for BC03 models \citep{mara11}. For the spectrum-fitting method in panel (c), the [Fe/H]$\rm_{S}$ of Vazdekis model is larger than that of the other two models, so the rms$\rm_{vazd}$ is smaller than rms$\rm_{bc03}$ and rms$\rm_{mara}$. From these rms, we can know that the dispersion of colour method is larger than that of other two methods. All these can not influence our conclusions, Lick-index method is suitable to study metallicity in the range of $-1.5\la$\,[Fe/H]\,$\la-0.7$ and spectrum-fitting method is suitable to study metallicity in the range of $-2.3\la$\,[Fe/H]\,$\la-1.5$.
From bottom panels, we can see that all these three methods have difficulties in age determinations for these two EPS models and the rms is larger than that of metallicity. There also exist some GCs with extreme large $\tau\rm_{I}$ (about 15.0\,Gyr) for these two models, and most of GCs have relative smaller $\tau\rm_{C}$, $\tau\rm_{I}$ and $\tau\rm_{S}$ than $\tau\rm_{L}$, which is similar to that of BC03 models. The comparisons of derived metallicities and ages of individual GCs show that these three models make different parameter  predictions. However,
the whole tendency for these three methods of these three EPS models is nearly the same. This indicates that our conclusions are independent of the EPS models.

\begin{figure}
\includegraphics[bb=95 45 590 690,height=8.0cm,width=6.0cm,clip,angle=270]{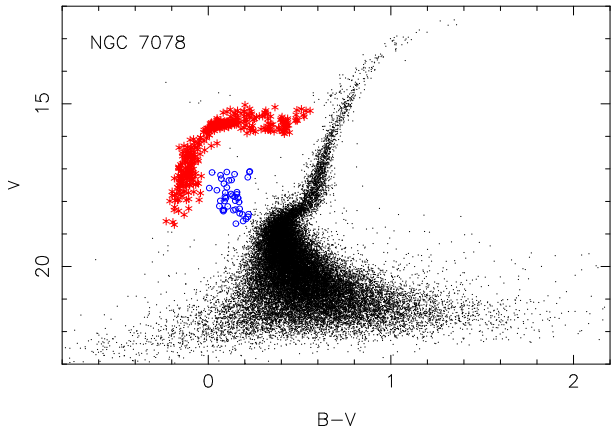}
\caption{Taking NGC 7078 as an example to illustrate the method of selecting HB stars and BSs from the CMD. The HB stars are marked with red asterisks and BSs are marked with blue circles.}\label{entropy-rlof}
\end{figure}

\begin{figure*}
\includegraphics[bb=35 15 555 780,height=16.cm,width=9.5cm,clip,angle=270]{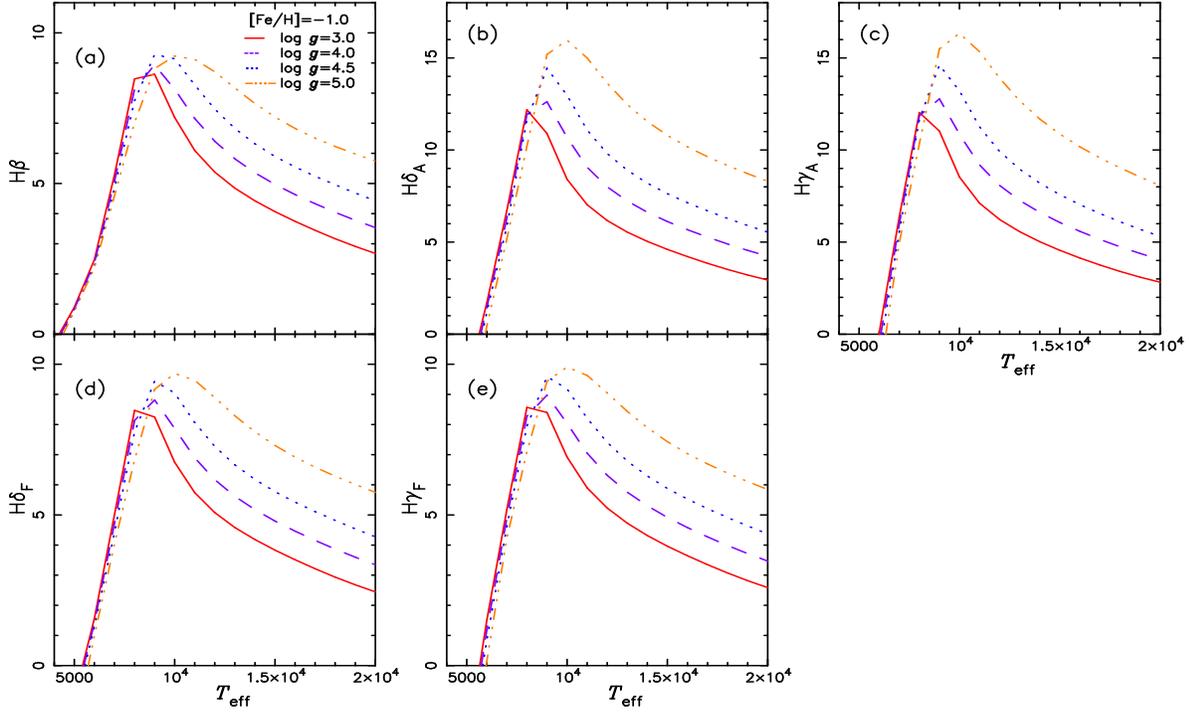} 
\caption{Theoretical Balmer index$-$\emph{T}$\rm_{eff}$ relations obtained from the B\tiny{LUERED} \normalsize{} library for different log \emph{g} (solid, dashed, dotted, and dash-dot-dot-dot lines stand for log \emph{g}$=3.0, 4.0, 4.5$ and $5.0$, respectively) at [Fe/H]$=-1.0$.}\label{entropy-rlof}
\end{figure*}
\section{Uncertainties in age}

From the above analyses, we know that our metallicities are consistent
with the values of literatures within the error bars, but there exist
some discrepancies in age determinations. It is well known that the HB stars, BSs, binary stars and $\alpha-$enhancement exist in GCs,
however they are not considered in most of the EPS models and they are important for models. All these can 
affect the parameter determinations for GCs.
So in this Section, we analyze the effects of some factors (HB morphology, BSs, binary star and $\alpha-$enhancement) on
age determinations.
\subsection{Horizontal Branch morphology}
It is widely known that HB morphology can affect
the age determination (de Freitas Pacheco \& Barbuy 1995; Lee et al. 2000; Maraston \& Thomas 2000). The HB stars
(especially the blue HB stars) can be hot and their existence can mimic young or
intermediate age populations in old SP systems. Hot extended HB stars have
been observed in GCs with HST by Piotto et al. (1999).

The main goal of this part is to analyze the contributions of
observed HB stars to the age determinations for three methods. At first,
we should select HB stars from the CMD and then give corresponding
colours, spectra and Lick indices for HB stars. We use
the distribution of stars in the CMD of P02 and select 18 GCs (include five with blue HBR,
four with unknown HBR and nine with red HBR), which have in common with our sample and show obvious
HB stars, to study the influence of HB stars on age determinations.
We select HB stars according to the method
described in Gratton et al. (2010), and take NGC 7078 as an example to show the selection of HB stars in Fig.\,6 
(red asterisks are HB stars).

The total observed quantity ($X\rm_{tot}$) of a
GC includes two components: the total HB stars
component (\emph{X}$\rm_{HB}$) and the component (\emph{X}$\rm_{(tot-HB)}$)
of other stars in the GC. After selecting of HB stars and analyzing $X\rm_{tot}$ and \emph{X}$\rm_{(tot-HB)}$,
we can study the influence of HB stars on age determinations.
The observed spectra and colours of these two components (\emph{X}$\rm_{HB}$ and \emph{X}$\rm_{(tot-HB)}$) can be constructed by follows.
\\(i) The integrated colours of these two components, taking $(B-V)\rm_{HB}$ and $(B-V)\rm\rm_{(tot-HB)}$ as examples, can be expressed as

\begin{equation}
\begin{array}{ll}
\;\;\;\;\;\;\emph{(B$-$V)}\rm_{HB}
=-2.5\,\log\frac{
\sum_{\tiny{\emph{i}}=1}^{\tiny{\emph{N}\rm_{HB}}}10^{-0.4\tiny{\emph{b}_{\emph{i}\rm,HB}}}}{\sum_{\tiny{\emph{i}}=1}^{\tiny{\emph{N}\rm_{HB}}}10^{-0.4\tiny{\emph{v}_{\emph{i}\rm,HB}}}}\;\;\;\;\rm{and}
\end{array}
\end{equation}
\begin{equation}
\begin{array}{ll}
(B-V)\rm_{(tot-HB)}
=-2.5\,\log\frac{10^{-0.4\tiny{\emph{B}}}-
\sum_{\tiny{\emph{i}}=1}^{\tiny{\emph{N}\rm_{HB}}}10^{-0.4\tiny{\emph{b}_{\emph{i}\rm,HB}}}}{10^{-0.4\tiny{\emph{V}}}-\sum_{\tiny{\emph{i}}=1}^{\tiny{\emph{N}\rm_{HB}}}10^{-0.4\tiny{\emph{v}_{\emph{i}\rm,HB}}}},
\end{array}
\end{equation}
where \emph{N}$\rm_{HB}$ is the number of HB stars,
\emph{b}$_{i\rm,HB}$ and \emph{v}$_{i\rm,HB}$ are the magnitudes for the $i$-th HB star which given by P02, and $B$ and $V$ are the magnitudes for each GC which obtained from Harris$'$ catalogue.
\\(ii) The spectra of \emph{F}$\rm_{HB}$ and \emph{F}$\rm_{(tot-HB)}$ can be obtained by following
\begin{equation}
\begin{array}{ll}
\;\;\;\;\;\;\;\;\;\;\;\;\;\emph{F}\rm_{HB}= \sum\limits_{\tiny{\emph{i}}=1}^{\tiny{\emph{N}}\rm_{HB}}\emph{f}\rm_{HB}^\emph{\,i} \;\;\;and
\end{array}
\end{equation}

\begin{equation}
\begin{array}{ll}

\;\;\;\;\;\;\;\;\;\;\;\;\;\emph{F}\rm_{(tot-HB)}= \emph{F}\rm_{tot} -\emph{F}\rm_{HB}\;,
\end{array}
\end{equation}
where $F$$\rm_{tot}$ is the spectrum taken from S05, and $\emph{f}^{ \;i}\rm_{HB}$ is the spectrum of $\emph{i}$\rm-th HB star which is obtained from the B\tiny{LUERED} \normalsize{library}. In this procedure, we transform the surface flux 
($\emph{f}^{ \;i}\rm_{sur}$) from B\tiny{LUERED} \normalsize{library} into observed flux ($\emph{f}^{ \;i}\rm_{HB}$) for each HB star by adopting the distance of GCs from the sun by Harris
catalogue.

For the colours and spectra of HB stars, we can derive them from the stellar spectral library based on  HB star$^{'}$s [Fe/H], \emph{T}$\rm_{eff}$ and
log\emph{g}. Because the GC is taken as a SSP, the same metallicity (i.e. [Fe/H]$\rm_{L}$ in Table A2) of GC
is adopted for HB stars for a certain GC. The  parameters of
\emph{T}$\rm_{eff}$ and log\emph{g} are obtained as follows.

(i) The \emph{T}$\rm_{eff}$ can be obtained from the theoretical
$BaSeL$ library by the $bv$ magnitudes of P02 and metallicities of GCs. From the theretical colour$-$\emph{T}$\rm_{eff}$ calibration of $BaSeL$ library, we can assign the \emph{T}$\rm_{eff}$ for each HB star.

(ii) The log\emph{g} is not easy to obtain, in this work we select a
bimodal log\emph{g} for HB stars, a median value of log\emph{g}\,$=$\,2.5 for HB stars with \emph{T}$\rm_{eff}$$\la$8000 K (include all red and part blue HB stars) and log\emph{g}\,$=$\,4.0 for HB stars with \emph{T}$\rm_{eff}$$\ga$8000 K (consist of blue HB stars). The reasons are given as follows.

(a) Moni Bidin et al. (2007) studied the \emph{T}$\rm_{eff}$, log\emph{g},
helium abundances and masses for HB stars in NGC 6752.
And they found that the log\emph{g} of all HB stars covered the range of
2.5\,$\leqslant$\,log\emph{g}\,$\leqslant$\,5.7. Fig.\,3 of \citet{mara03} displayed the theoretical isochrones, 
and the log\emph{g} range for HB stars was about $2.0-4.0$. \citet{dorm92} and \citet[in preparation]{lei11} have presented a large grid of HB evolution sequences (including the evolution of extreme HB stars), and from their data we find the log\emph{g} range for red HB star is about $1.8-3.0$ and for blue HB star is about $3.0-5.5$.

(b) The Balmer indices have a strong dependence on the log\emph{g} for \emph{T}$\rm_{eff}$$\ga$8000 K,
they show decrease with increasing  \emph{T}$\rm_{eff}$ within a fixed metallicity. This is shown in Fig.\,7, from it we see that the differences are large for different log\emph{g} with \emph{T}$\rm_{eff}$$\ga$8000 K. So for the HB stars with \emph{T}$\rm_{eff}$$\la$8000 K, the adoption of log\emph{g} has little effect on Balmer indices, but for those with \emph{T}$\rm_{eff}$$\ga$8000 K, the adoption of log\emph{g} should be reasonable. In addition,  from Fig.\,1 of \citet{reci06}, we can see that all red and part blue HB stars are with \emph{T}$\rm_{eff}$$\la$8000 K and part blue HB stars are with \emph{T}$\rm_{eff}$$\ga$8000 K. Combined with log\emph{g} range of red and blue HB stars, we choose a bimodal log$g$ for them and the boundary is \emph{T}$\rm_{eff}$$\approx$8000 K, log\emph{g}\,$=$\,2.5 (the median value of red HB stars) for those with \emph{T}$\rm_{eff}$$\la$8000 K and log\emph{g}\,$=$\,4.0 (the median value of blue HB stars) for those with \emph{T}$\rm_{eff}$$\ga$8000 K.

\begin{figure*}
\includegraphics[bb=30 20 580 770,height=16.0cm,width=9.5cm,clip,angle=270]{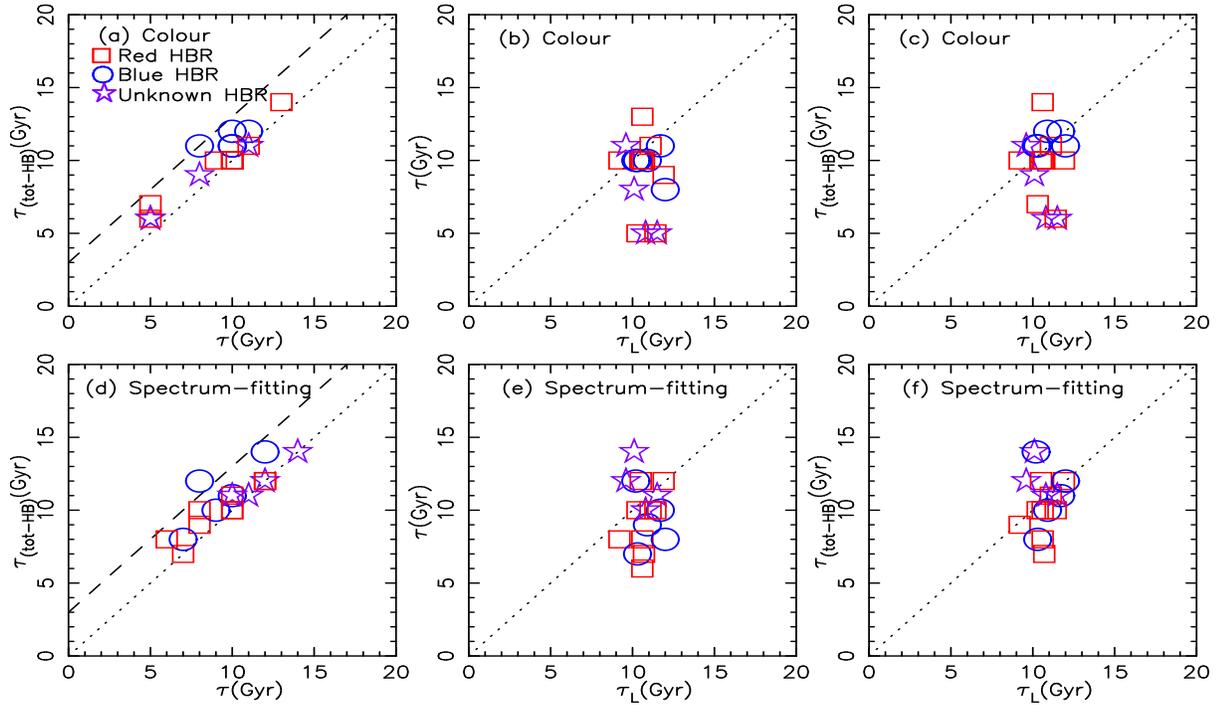}
\caption{The influence of HB stars on the age determinations for colour (top panels) and spectrum-fitting (bottom panels) methods. Red open squares, blue open circles and green open pentacles stand for GCs with red, blue and unknown HBR, respectively.
The dotted lines present 1:1 relation, the dashed lines indicate the range for $\tau\rm_{(tot-HB)}$$-\tau=3.0$. }\label{entropy-rlof}

\end{figure*}
\begin{figure*}
\includegraphics[bb=25 20 590 770,height=16.0cm,width=9.5cm,clip,angle=270]{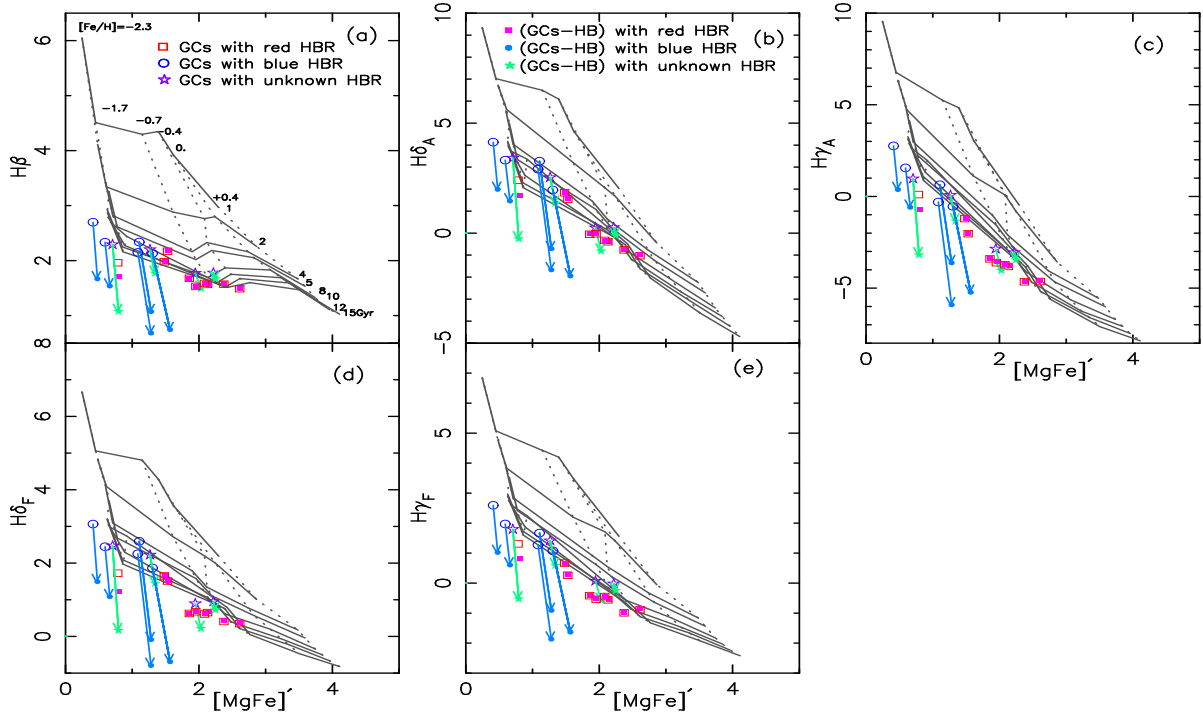}
\caption{The influence of HB stars on the age determinations for Lick-index method. Lines are the same as in Fig.\,3. The solid symbols are the GCs with HB stars and the open symbols are the GCs without HB stars. The squares, circles and pentacles stand for GCs with red, blue and unknown HBR, respectively. The arrows indicate how the indices for GCs change when HB stars are removed, and we do not show the arrow for GCs with red HBR because the influence of HB stars for this kind of GC is very small. }\label{entropy-rlof}

\end{figure*}

Based on [Fe/H], \emph{T}$\rm_{eff}$ and log\emph{g} determined as described above and the stellar spectral library,
we can obtain the \emph{F}$\rm_{HB}$ and \emph{F}$\rm_{(tot-HB)}$. Using the methods described in Section 3 we can study the influence of HB stars on age determinations. 

For the colour and spectrum-fitting methods, we show
the influences of HB stars on age determinations in Fig.\,8. For the sake of simplicity, $\tau$ represents the age of GC with HB stars (correspond to $\tau\rm_{C}$ and $\tau\rm_{S}$ in Section 4.2) and $\tau\rm_{(tot-HB)}$ represents the age of GC without HB stars. We fix [Fe/H]$\rm_{C}$ and [Fe/H]$\rm_{S}$ for GCs when calculating $\tau\rm_{(tot-HB)}$. For colour method, panel (a) shows the comparison between $\tau\rm_{(tot-HB)}$ and $\tau$. And panels (b) and (c) represent the level of comparisons between a case of ($\tau\rm_{L}$, $\tau$) and a case of ($\tau\rm_{L}$, $\tau\rm_{(tot-HB)}$). We give the range for $\tau\rm_{(tot-HB)}-$$\tau=3.0$ by dashed line in panel (a). We can see all of GCs with blue HBR are with $\tau\rm_{(tot-HB)}$$>$$\tau$, and part of GCs with red and unknown HBR are with $\tau\rm_{(tot-HB)}$$>$$\tau$. On the whole, the influence of HB stars on age determination is larger for GCs with blue HBR than those with red and unknown HBR. From panels (b) and (c), we can see that most of GCs are on the $\tau\leq$$\tau\rm_{L}$ side in panel (b), while the GCs are on both side of the dotted line (1:1 relation, $\tau\rm_{(tot-HB)}$$=\tau\rm_{L}$) in panel (c), and the dispersion between $\tau\rm_{(tot-HB)}$ and $\tau\rm_{L}$ is smaller than that between $\tau$ and $\tau\rm_{L}$. So the consistence between $\tau\rm_{(tot-HB)}$ and $\tau\rm_{L}$ is better than that between $\tau$ and $\tau\rm_{L}$ excepting four GCs with small $\tau\rm_{(tot-HB)}$ and $\tau$. These four GCs are with $\tau=5.0$, and they are the GCs (NGC 6342, 6338, 6441 and 6652) said in Section\,4.2. From these two panels we can know that the HB stars can change the age of these four GCs about 1.0$-$2.0\,Gyr and can not change age largely. Because we fix [Fe/H]$\rm_{C}$ on calculating $\tau\rm_{(tot-HB)}$ and two (NGC 6342 and 6652) of these four GCs have relative large [Fe/H]$\rm_{C}$, these two GCs can be affected by the age-metallicity degeneracy. And the result would be $\tau\rm_{(tot-HB)}$$\gg$$\tau$ if [Fe/H]$\rm_{C}$ is not fixed. However, for the other two GCs, they have a bit larger [Fe/H]$\rm_{C}$, excepting affected by age-metallicity degeneracy, they also can be affected by their positions on Galactic plane for they being close to Galactic bulge.

For spectrum-fitting method, just as colour method, panels (d), (e) and (f) show the comparisons between $\tau\rm_{(tot-HB)}$, $\tau$ and $\tau\rm_{L}$. And we give the range for $\tau\rm_{(tot-HB)}$$-\tau=3.0$ by dashed line in panel (d). We can see all of GCs with blue HBR are with $\tau\rm_{(tot-HB)}$$>$$\tau$, and part of GCs with red and unknown HBR are with $\tau\rm_{(tot-HB)}$$>$$\tau$. On the whole, the influence of HB stars on age determination is larger for GCs with blue HBR than those with red and unknown HBR. From panels (e) and (f), we also can see that the dispersion between $\tau\rm_{(tot-HB)}$ and $\tau\rm_{L}$ is smaller than that between $\tau$ and $\tau\rm_{L}$ and the GCs is more close to the dotted line (1:1 relation) in panel (e), so the consistence between $\tau\rm_{(tot-HB)}$ and $\tau\rm_{L}$ is better than that between $\tau$ and $\tau\rm_{L}$.

In the 10th and 12th columns of Table A2, we give the change of age ($\Delta\tau$\tiny{$\rm^{C}_{HB}$ \normalsize{} and $\Delta\tau$\tiny{$\rm^{S}_{HB}$ \normalsize{}, $\tau\rm_{(tot-HB)}-\tau$) influenced by HB stars for these two methods. 
On the whole, the existence of HB stars can make age be small about 0.0$-$3.0\,Gyr for this two methods. 
\begin{figure*}
\includegraphics[bb=30 20 580 770,height=16.0cm,width=9.5cm,clip,angle=270]{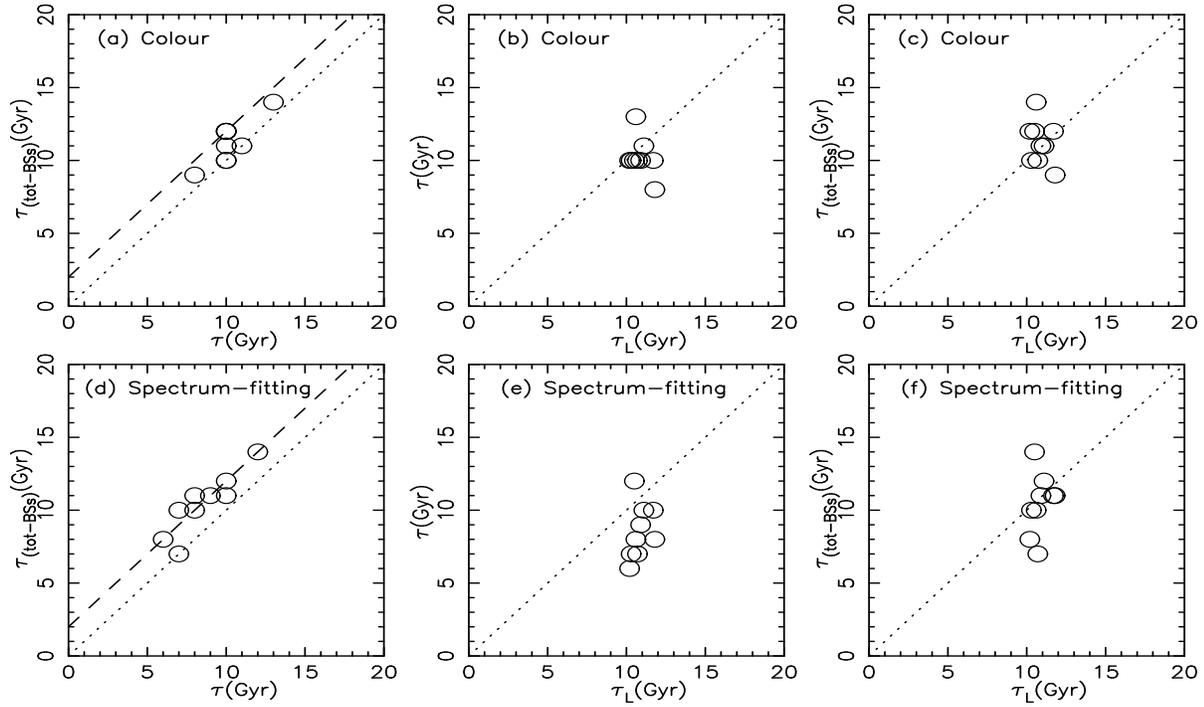}
\caption{The same as Fig.\,8, but for the influence of BSs on the age determinations.}\label{entropy-rlof}

\end{figure*}
\begin{figure*}
\includegraphics[bb=25 20 590 770,height=16.0cm,width=9.5cm,clip,angle=270]{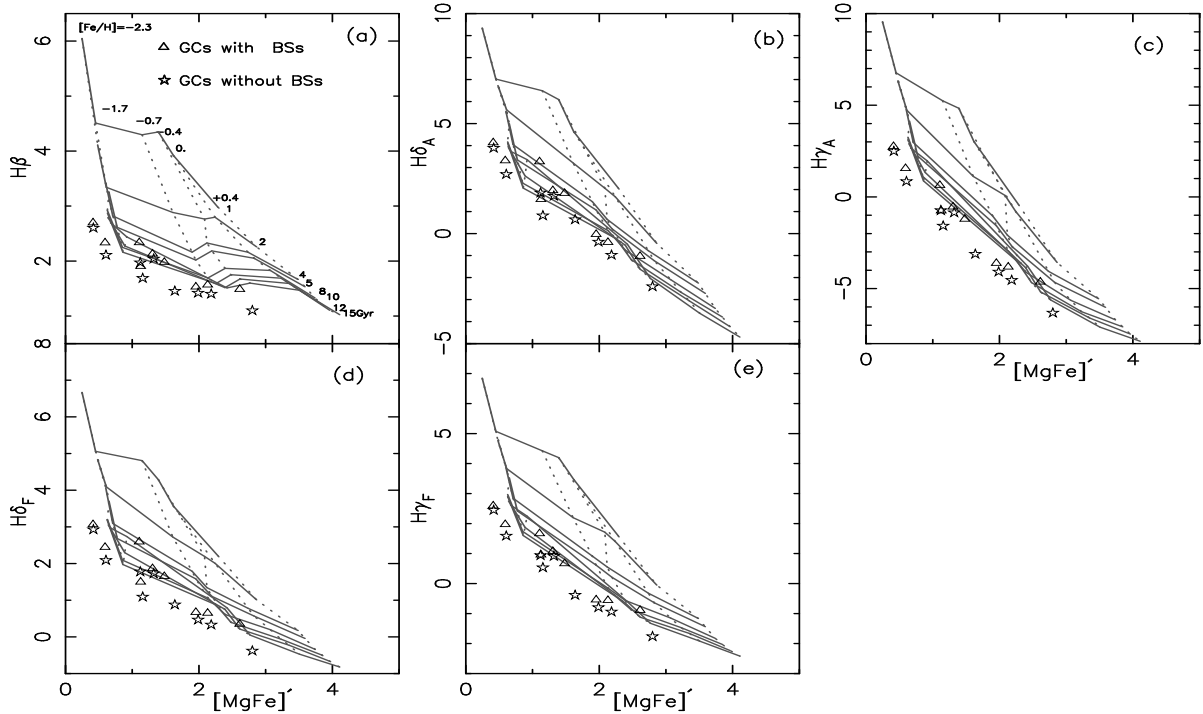}
\caption{The influence of BSs on the age determinations of GCs for Lick-index method. Lines are the same as in Fig.\,9. Open triangles are the GCs with BSs, and open pentacles are the GCs without BSs.  }\label{entropy-rlof}

\end{figure*}

For the Lick-index method, the influence is displayed in Fig.\,9, in which the open and solid symbols stand for GCs with and without HB stars, respectively; squares, circles and pentacles display GCs with red, blue and unknown HBR, respectively. The positions of GCs with and without HB stars change a lot for those with blue and unknown HBR and small for those with red HBR. For clarity, we use arrows to indicate the moving directions when HB stars removed for GCs with blue and unknown HBR whose positions change large, and do not use arrows to indicate the moving directions for GCs with red HBR whose positions change small. From these arrows, we can see that the positions for those with blue and unknown HBR are shifted toward older age ($>$15.0\,Gyr). All these show that the ages of GCs with blue and unknown HBR can be affected strongly by HB stars, and the ages of those with red HBR are affected slightly by HB stars. In the 11th column of Table A2, we list the change of age ($\Delta\tau$\tiny{$\rm^{I}_{HB}$ \normalsize{}, $\tau\rm_{(tot-HB)}-\tau$) affected by HB stars for Lick-index method. For the GCs with and without HB stars all lying inside Lick indices grids, we give the numerical change of age. For those with HB stars lying inside grids and lying beyond grids when HB stars removed, we assume $\Delta\tau\geqslant5.0$\,Gyr to show the change of age even though this may be just the minimal change of age. For those with and without HB stars all lying beyond grids, we use $'$uncom$'$ instead of numerical value to express the change of age, but this does not mean that the age of GCs changes small. On the whole, the lower limit of maximal
change of age is 6.0\,Gyr due to the influence of HB stars, and the influence of HB stars is stronger for GCs with blue and unknown HBR than those with red HBR.

\subsection{Blue straggler stars}
BSs are identified as blue, luminous extensions of main-sequence (MS) stars
(Sandage 1953), and they have
been widely observed in stellar systems (Johnson et al. 1999; Alcaino et al. 2003; Piotto et al. 2004;
Ahumada \& Lapasset 2007).
Just as the blue HB stars, the existence of
BSs can also mimic the presence of younger SPs (Lee et al. 2000;
Schiavon et al. 2004; Cenarro et al. 2008). They can enhance
the integrated spectrum in ultraviolet and U bands of the GC. This phenomenon would lead to
the GC being predicted as young or lower metallicity based on
the EPS models without considering the BSs. Cenarro et al. (2008)
found that higher BSs ratios can lead to smaller
apparent ages. Xin et al. (2008) studied the BSs$^{,}$ influence on the
integrated properties of Large Megellanic Cloud star cluster ESO
121\,$-$\,SC03, and found that the
best-fitting values of age and metallicity are significantly underestimated compared to the true
cluster parameters.

Similar to the procedure of investigating the influence of HB stars on age determinations in Section 5.2, we also calculate the influence of BSs on age determinations.
The first step, we select the BSs from the CMDs of P02 through directly visual inspection of the CMD of each cluster (Fig.\,6, just as the method describing in Moretti, Angeli $\&$ Piotto 2008). And in this work we select 9 GCs (in common with our sample) with distinct BSs to study the quantitative influence of BSs. The parameter of \emph{T}$\rm_{eff}$ for each BSs can be estimated by the empirical relation
$\log$\emph{T}$\rm_{eff}$$ = -0.38 (b\rm_{BS}-\emph{v}\rm_{BS}) + 3.99$ \citep{ferr06}. Ferraro et al. (2006) also indicated that the gravity of BSs was in the range of 4.3$\leq$\,$\log$\emph{g}\,$\leq$4.8.
In our procedure, we choose $\log$\emph{g}\,=\,4.5 for BSs.
The same metallicity of GC is adopted for BSs of a certain GC. Based on these parameters we can obtain the colours, Lick indices and spectra of BSs from stellar spectral library. Because the B\tiny{LUERED} \normalsize{library} is at high resolution, it does not provide colours. We adopt the colours from $BaSeL$ library and spectra from B\tiny{LUERED} \normalsize{library}.

Using the methods described in Section\,3, we can study the influence of BSs on age determinations. Similar to Fig.\,8,
Fig.\,10 shows the influence of BSs on age determinations for colour and spectrum-fitting methods, and $\tau\rm_{(tot-BSs)}$ represents the age of GC without BSs. For colour method, panles (a), (b) and (c) show the comparisons between $\tau\rm_{(tot-BSs)}$, $\tau$ and $\tau\rm_{L}$. We give the range for $\tau\rm_{(tot-BSs)}-$$\tau=2.0$ by dashed line in panel (a). We can see that most GCs are with $\tau\rm_{(tot-BSs)}$$>$$\tau$, and this indicates that the existence of BSs can make GCs look younger. From panels (b) and (c), we can see the GCs lie on both side of the dotted line in panel (c) and lie on one side of the dotted line in panel (b), so the consistence between $\tau\rm_{(tot-BSs)}$ and $\tau\rm_{L}$ is better than that between $\tau$ and $\tau\rm_{L}$.

For spectrum-fitting method, similar to colour method, panels (d), (e) and (f) show the comparisons between $\tau\rm_{(tot-BSs)}$, $\tau$ and $\tau\rm_{L}$. And we give the range for $\tau\rm_{(tot-BSs)}-$$\tau=2.0$ by dashed line in panel (d). Just as the colour method, most GCs are with $\tau\rm_{(tot-BSs)}$$>$$\tau$. From panels (e) and (f), we can see that the dispersion between $\tau\rm_{(tot-BSs)}$ and $\tau\rm_{L}$ is smaller than that between $\tau$ and $\tau\rm_{L}$, and the GCs is more close to the dotted line (1:1 relation) in panel (e), so the consistence between $\tau\rm_{(tot-BSs)}$ and $\tau\rm_{L}$ is better than that between $\tau$ and $\tau\rm_{L}$. 

In the 13th and 15th columns of Table A2, we give the change of age ($\Delta\tau$\tiny{$\rm^{C}_{BSs}$ \normalsize{} and $\Delta\tau$\tiny{$\rm^{S}_{BSs}$ \normalsize{}, $\tau\rm_{(tot-BSs)}-\tau$) influenced by BSs for these two methods. The change of age is about 0.0$-$2.0\,Gyr for the influence of BSs, and this approximates to influence (0.0$-$3.0\,Gyr) of HB stars on age determinations for these two methods. In theory, the \emph{T}$\rm_{eff}$ of HB stars is higher than BSs, and the existence of HB stars can make the spectra and
colours bluer than that of BSs. But these two type stars have great contribution in blue and ultraviolet bands. In this work, we select $UBVI$ colours and the spectra in the 3700$-$5700\,${\rm\AA}$ range, so the difference of influences by these two type stars is small in such range of colours and spectra, and the change of age affected by these two type stars is nearly the same. The difference of influences on age by these two type stars may be obvious for Lick-index method.

 Fig.\,11 displays the effect of BSs on age determination for Lick-index method. Open triangles and pentacles stand for GCs with and without BSs, respectively. From these different symbols, we can see that the Balmer indices of GCs with BSs are stronger than those without BSs, and those without BSs go down on the Lick indices planes. Similar to HB stars, the change of age ($\Delta\tau$\tiny{$\rm^{C}_{BSs}$ \normalsize{}, $\tau\rm_{(tot-BSs)}-\tau$) affected by BSs for Lick-index method is listed in the 14th column of Table A2. On the whole, the lower limit of maximal change of age is 5.0\,Gyr due to the existence of BSs, and this approximates to the lower limit of maximal change of age affected by HB stars (6.0\,Gyr). We can not say that the influences of this two type stars are similar, because we do not give the numerical value of age change for those lying beyond grids. From Figs.\,9 and 11, we can see that HB stars can make H$\beta$ stronger about 1.5$\rm\AA$, but BSs can make H$\beta$ stronger about 0.5$\rm\AA$, this demonstrate that the influence on age determination for HB stars is larger than that of BSs. And the difference of influence on age by these two type stars is obvious for this method.

\begin{figure}
\includegraphics[bb=28 20 590 770,height=8.cm,width=5.0cm,clip,angle=270]{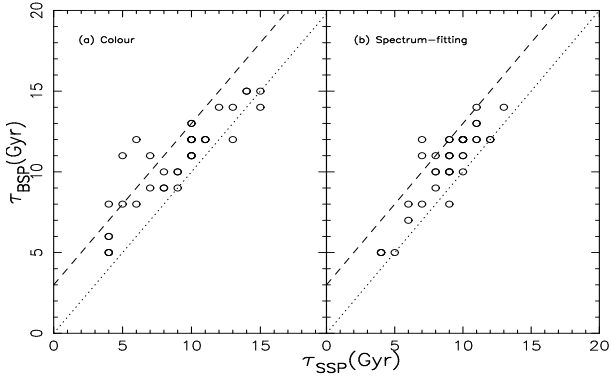}
\caption{The influence of binary stars on age determinations for colour (left panel) and spectrum-fitting (right panel) methods.
The dotted and dashed lines have the same meaning with Fig.\,8. }\label{entropy-rlof}

\end{figure}
\begin{figure*}
\includegraphics[bb=25 20 590 770,height=16.0cm,width=9.5cm,clip,angle=270]{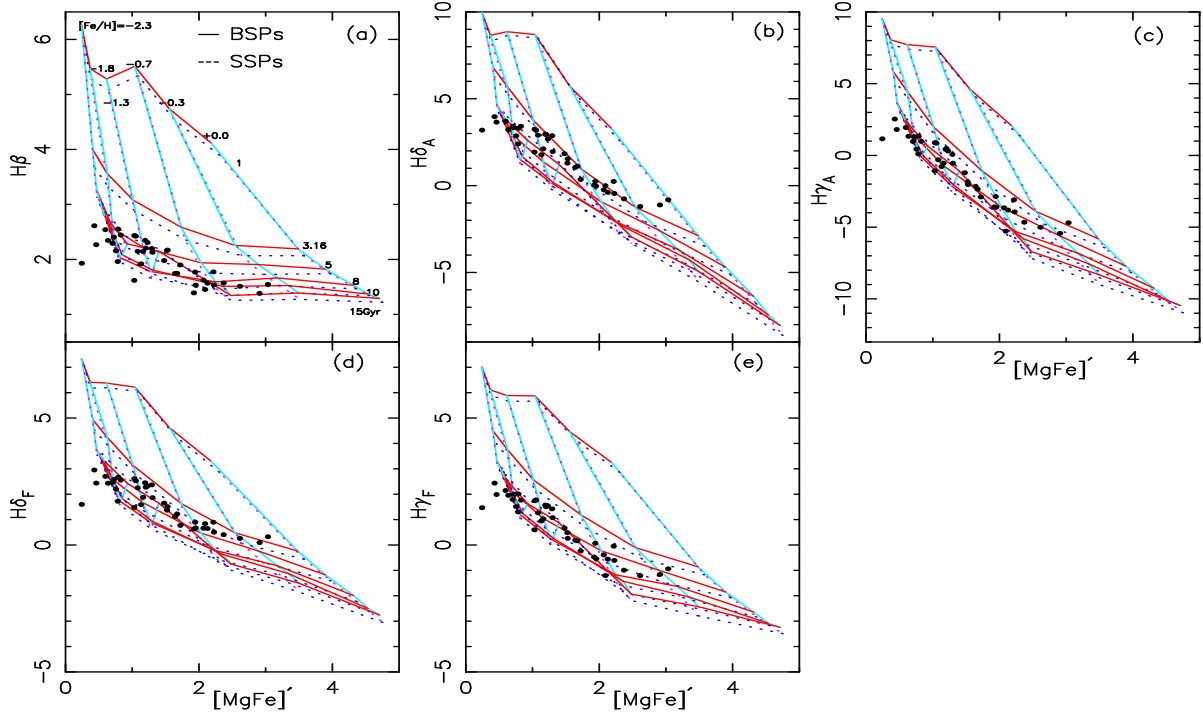}
\caption{The influence of binary interactions on the age determinations on Balmer vs. [MgFe]$^{'}$ panels for some GCs. Solid and dashed lines stand for BSP and SSP models, respectively. The dots stand for GCs.  }\label{entropy-rlof}

\end{figure*}

\subsection{Binary stars}

Binary stars are very common in star clusters and galaxies. The
evolution of binary stars is very different from single stars, and
binary interactions can also create some important objects and
phenomenons, such as BSs (e.g. Pols \& Marinus 1994; Tian et al.
2006), subdwarf B stars (Han et al. 2002, 2003), and the
Gravitational wave radiation sources (Liu 2009). Therefore, binary
stars have the potential to play an important role in determining
the overall appearance of any realistic SP. Zhang et
al. (2004, 2005a) have investigated the effects of binary
interactions on the integrated colours, spectra and Lick indies. And they found that the
inclusion of binary interactions made the integrated $U-B$ and
$B-V$ colours and spectra of populations bluer for various metallicities, and
made the H$\beta$ greater than that without binary interactions.

In this work, we investigate the effect of binary stars on the age determinations of GCs based on the models of SSPs (Zhang et al. 2004) and BSPs (Zhang et al. 2005a). For these two models, we adopt the B\tiny{LUERED}\normalsize{} library instead of $BaSeL$ for spectrum-fitting and Lick-index methods, and $BaSeL$ library for colour method. More description of these two models is given in Section 2.2.
In this part we obtain the ages of GCs from three methods based on SSP and BSP models.  In Fig.\,12 we display the ages determined from SSP and BSP models for colour and spectrum-fitting methods. The ages of GCs obtained from BSP model are lager than those obtained from SSP model. Similar to the HB and BS stars, the existence of binary star can make the GCs look younger (about $0.0-3.0$\,Gyr) for these two methods. Similar to above, binary stars also have great contributions in blue and ultraviolet bands. So the influence of binary stars is small in the colours and spectra that we used. Fig.\,13 gives the results for Lick-index method, in which the solid and dashed lines stand for BSP and SSP models, respectively. From them we can see that the grids for BSP model go up in these five panels, and the five  Balmer indices of BSP model are greater by $\sim$0.15 ${\rm\AA}$ than those of SSP model. Using these two models to study the ages for GCs, we can see that the lower limit of maximal change of age is 3.0\,Gyr for the existence of binary stars.
\begin{figure*}
\includegraphics[bb=25 20 590 760,height=16.cm,width=10.0cm,clip,angle=270]{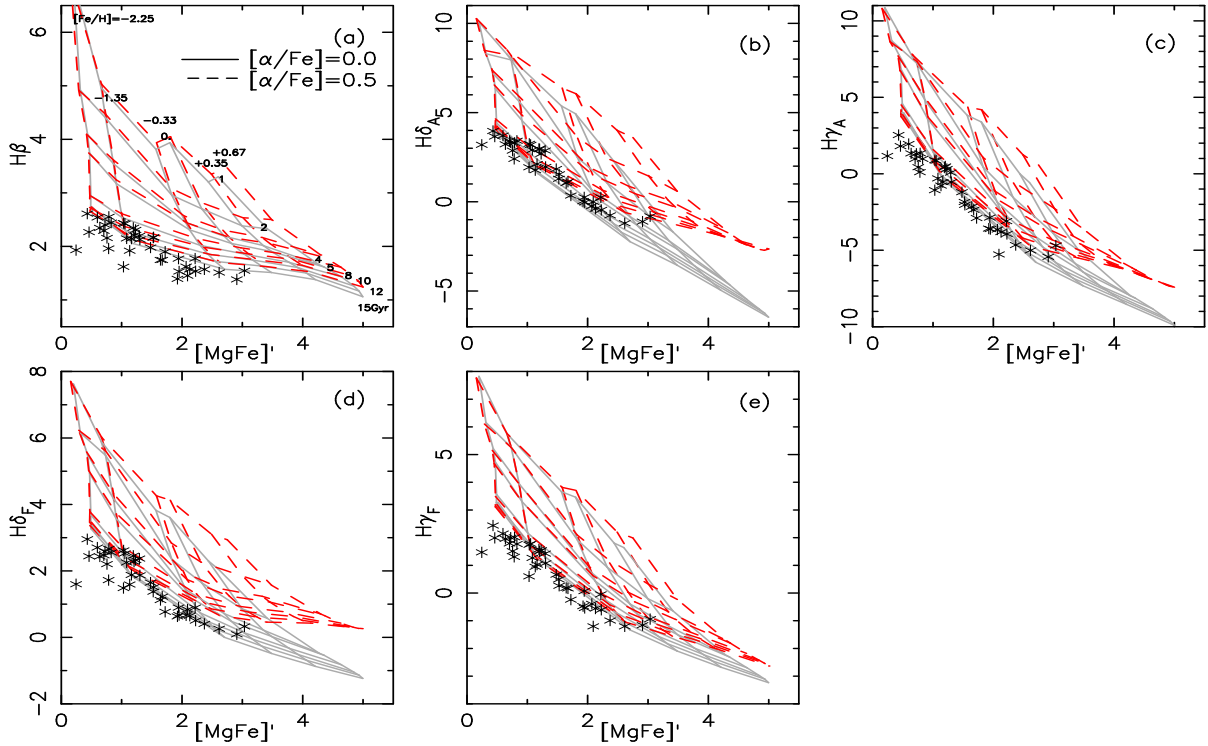}
\caption{The distribution of GCs on the Balmer indices and [MgFe]$^{'}$ planes based on 
\citet{thom11a} model. The lines are the same as in Fig.\,2, the gray solid and red dashed lines stand for [$\alpha$/Fe]$=0.0$ and $0.5$, respectively.}\label{entropy-rlof}

\end{figure*}

\subsection{$\alpha-$enhancement}

The existence of $\alpha-$enhancement can make stars on the isochrones hotter throughout all evolutionary phases, and the effect increases with metallicity. Galactic GCs are found to be $\alpha-$enhanced \citep{thom03,mara03,lee05}. Just as the other factors, the existence of $\alpha-$enhancement also can mimic young SPs and make Galactic GCs look younger. For the Lick-index method, as said in Section\,3.2, the Balmer indices would be affected by $\alpha-$enhancement, so the age determined by Lick-index method can be affected by this factor. \citet{mara11} and \citet{thom11b} have selected some Lick indices and used the $\chi$$^{2}-$fitting technique to derive the parameters for GCs, and they found that the derived ages agreed well with those obtained by CMDs when taking account the $\alpha-$enhancement. \citet{thom03,thom11a} have presented the whole set of Lick indices of SSP models with variable $\alpha$/Fe ratio, this allows us to study the influence of $\alpha-$enhancement on age determination for Lick-index method.

We use the Lick indices of \citet{thom11a} models to study the influence of $\alpha-$enhancement on age determination for Lick-index method, and two types of $\alpha$/Fe ratio ([$\alpha$/Fe]$=0.0$ and [$\alpha$/Fe]$=+0.5$) are chosen. Fig.14 gives the result, from them we can see that $\alpha-$enhancement can make the Balmer indices stronger, especial for high metallicity, and has little effect on [MgFe]$'$ which consists with above said (Section 3.2). Therefore the $\alpha-$enhancement can affect the age determination and has little effect on metallicity determination for Lick-index method. We also can see that the influence of $\alpha-$enhancement on H$\beta$ and H$\gamma_{F}$ is small at the range of [Fe/H]\,$\la-0.7$. In this work, the ages for GCs are computed as the weighted mean of the ages derived from five Balmer indices versus [MgFe]$'$ grids and most of the GCs are in the range of [Fe/H]\,$\la-0.7$. Therefore, the $\alpha-$enhancement has little effect on age determination of GCs for Lick-index method. But for the study of high-metallicity SP systems, the $\alpha-$enhancement should be taken into account, because the influences of $\alpha-$enhancement on these five Balmer indices are all significant for high-metallicity range.

\section{Summary}
We have investigated the utility of three methods (colour,
Lick-index and full spectrum-fitting) with the same EPS models
(BC03 models) for estimating ages and metallicities of
Galactic GCs. We also compared our results with those
estimated in the literatures from other methods, and the
main results of this study are as follows.

(1) Our results show that Galactic GCs are almost old metal-poor
SP systems.

(2) Metallicities determined from these three methods are in agreement with
those of literatures in the entire range spanned by GCs.
The Lick-index method is suitable to obtain 
metallicity for SP systems in the range of $-1.5\la$[Fe/H]$\la-0.7$ and spectrum-fitting method is suitable to study metallicity for SP systems in the range of $-2.3\la$[Fe/H]$\la-1.5$.

(3) There exist some discrepancies between our ages and the literatures, we can not directly
say which method is more suitable for age determination.
Through comparing with the literatures we find our results
are younger than the literatures on average, especially for GCs with
blue HBR. Many factors can affect the age determinations and make GCs become younger (such as the HB morphology,
BSs, binary stars, $\alpha-$enhancement).

(4) We use Vazdekis and Maraston models to investigate whether our results are dependent on EPS models. The comparisons of derived metallicities and ages for individual GCs show that three models make different parameter predictions. However,
the whole tendency for these three methods of these three EPS models is nearly the same. All these indicate that our above conclusions are independent of the EPS models.

(5) In this work we use the old Galactic GCs to test these three methods, and our results may hold for old SP system.

(6) We also study the quantitative influences of HB, BSs and binary stars on age determinations for three methods.
The existence of all these stars can make the GCs look younger. For the colour and spectrum-fitting methods, the age can be underestimated about 0.0$-$3.0\,Gyr, 0.0$-$2.0\,Gyr and 0.0$-$3.0\,Gyr due to influences of HB, BSs and binary stars, respectively.
And for Lick-index method, the lower limit of maximal change of age is 6.0\,Gyr, 5.0\,Gyr and 3.0\,Gyr due to influences of horizontal branch, blue straggler and binary stars, respectively.
 
(7) For the Lick-index method, we also investigate the influence of $\alpha-$enhancement
on age determination, and find that the $\alpha-$enhancement has little effect on age determination for Galactic GCs.



\section*{Acknowledgments}
We thank Richard de Grijis for some useful discussions and suggestions. We thank Claudia Maraston 
for kindly providing the new version of Maraston EPS model. We thank the anonymous referee for very 
valuable comments that help to improve the manuscript.
This work is in part supported by Natural Science Foundation (Grant
Nos 10773026, 11033008, 10821061, 2007CB815406 and 11103054) and by the Chinese Academy
of Sciences under Grant No. KJCX2-YW-T24. The authors are also supported by the program of
the Light in China$^,$ Western Region (LCWR) under grant XBBS2011022, the Beyond The Horizons under grant 100020101 and
Xinjiang Natural Science Foundation (No.2011211A104).

\appendix

\section[]{The parameters of Galactic GCs and some influence on age determination}

\begin{table*}
\begin{center}
\caption[]{The main basic parameters for GCs. The first column gives the names for Galactic GCs. The 2nd and 3rd columns are right ascension (RA) and declination (DEC) (epoch J2000). Galactic longitude (L) and latitude (B) in degrees are listed in the 4th and 5th columns. Distance 
from Sun and Galactic center are shown in the 6th and 7th columns. The 8th column gives the reddening value, $E(B-V)$. The parameter describing the HB morphology is listed in the 9th column. All of these data are adopted from Harris$'$ catalogue.}
\tabcolsep=0.09in

\label{Tab:frac}
\renewcommand{\arraystretch}{1.4}

\begin{tabular}{|c|c|c|r|r|r|r|r|r|r|c|c|c|c|c|c|c|c|c|c|}
\hline\noalign{\smallskip}

(1) &(2)  &(3)  &(4)  &(5)  &(6)  &(7)  &(8)  &(9)\\
NGC  &RA   &DEC &L&B &\emph{R}\tiny{$\rm_{Sun}$} &\emph{R}\tiny{$\rm_{Gc}$}&$E(B-V)$ &HBR \\
          & (2000)   &(2000) &(deg) &(deg)  &(kpc) &(kpc)  \\

\hline\noalign{\smallskip}
104   &$00\;24\;05.67$&  $-72\;04\;52.6$&  $ 305.89$&  $-44.89$&  $  4.50$&  $  7.40$&  $  0.04$&  $ -0.99$\\
1851  &$05\;14\;06.76$&  $-40\;02\;47.6$&  $ 244.51$&  $-35.03$&  $ 12.10$&  $ 16.60$&  $  0.02$&  $ -0.36$\\
1904  &$05\;24\;11.09$&  $-24\;31\;29.0$&  $ 227.23$&  $-29.35$&  $ 12.90$&  $ 18.80$&  $  0.01$&  $  0.89$\\
2298  &$06\;48\;59.41$&  $-36\;00\;19.1$&  $ 245.63$&  $-16.00$&  $ 10.80$&  $ 15.80$&  $  0.14$&  $  0.93$\\
2808  &$09\;12\;03.10$&  $-64\;51\;48.6$&  $ 282.19$&  $-11.25$&  $  9.60$&  $ 11.10$&  $  0.22$&  $ -0.49$\\
3201  &$10\;17\;36.82$&  $-46\;24\;44.9$&  $ 277.23$&  $  8.64$&  $  4.90$&  $  8.80$&  $  0.24$&  $  0.08$\\
5286  &$13\;46\;26.81$&  $-51\;22\;27.3$&  $ 311.61$&  $ 10.57$&  $ 11.70$&  $  8.90$&  $  0.24$&  $  0.80$\\
5904  &$15\;18\;33.22$&  $+02\;04\;51.7$&  $   3.86$&  $ 46.80$&  $  7.50$&  $  6.20$&  $  0.03$&  $  0.31$\\
5927  &$15\;28\;00.69$&  $-50\;40\;22.9$&  $ 326.60$&  $  4.86$&  $  7.70$&  $  4.60$&  $  0.45$&  $ -1.00$\\
5946  &$15\;35\;28.52$&  $-50\;39\;34.8$&  $ 327.58$&  $  4.19$&  $ 10.60$&  $  5.80$&  $  0.54$&  ...     \\
5986  &$15\;46\;03.00$&  $-37\;47\;11.1$&  $ 337.02$&  $ 13.27$&  $ 10.40$&  $  4.80$&  $  0.28$&  $  0.97$\\
6121  &$16\;23\;35.22$&  $-26\;31\;32.7$&  $ 350.97$&  $ 15.97$&  $  2.20$&  $  5.90$&  $  0.35$&  $ -0.06$\\
6171  &$16\;32\;31.86$&  $-13\;03\;13.6$&  $   3.37$&  $ 23.01$&  $  6.40$&  $  3.30$&  $  0.33$&  $ -0.73$\\
6218  &$16\;47\;14.18$&  $-01\;56\;54.7$&  $  15.72$&  $ 26.31$&  $  4.80$&  $  4.50$&  $  0.19$&  $  0.97$\\
6235  &$16\;53\;25.31$&  $-22\;10\;38.8$&  $ 358.92$&  $ 13.52$&  $ 11.50$&  $  4.20$&  $  0.31$&  $  0.89$\\
6254  &$16\;57\;09.05$&  $-04\;06\;01.1$&  $  15.14$&  $ 23.08$&  $  4.40$&  $  4.60$&  $  0.28$&  $  0.98$\\
6266  &$17\;01\;12.80$&  $-30\;06\;49.4$&  $ 353.57$&  $  7.32$&  $  6.80$&  $  1.70$&  $  0.47$&  $  0.32$\\
6284  &$17\;04\;28.51$&  $-24\;45\;53.5$&  $ 358.35$&  $  9.94$&  $ 15.30$&  $  7.50$&  $  0.28$&  ...     \\
6304  &$17\;14\;32.25$&  $-29\;27\;43.3$&  $ 355.83$&  $  5.38$&  $  5.90$&  $  2.30$&  $  0.54$&  $ -1.00$\\
6316  &$17\;16\;37.30$&  $-28\;08\;24.4$&  $ 357.18$&  $  5.76$&  $ 10.40$&  $  2.60$&  $  0.54$&  $ -1.00$\\
6333  &$17\;19\;11.26$&  $-18\;30\;57.4$&  $   5.54$&  $ 10.71$&  $  7.90$&  $  1.70$&  $  0.38$&  $  0.87$\\
6342  &$17\;21\;10.08$&  $-19\;35\;14.7$&  $   4.90$&  $  9.72$&  $  8.50$&  $  1.70$&  $  0.46$&  $ -1.00$\\
6352  &$17\;25\;29.11$&  $-48\;25\;19.8$&  $ 341.42$&  $ -7.17$&  $  5.60$&  $  3.30$&  $  0.22$&  $ -1.00$\\
6356  &$17\;23\;34.93$&  $-17\;48\;46.9$&  $   6.72$&  $ 10.22$&  $ 15.10$&  $  7.50$&  $  0.28$&  $ -1.00$\\
6362  &$17\;31\;54.99$&  $-67\;02\;54.0$&  $ 325.55$&  $-17.57$&  $  7.60$&  $  5.10$&  $  0.09$&  $ -0.58$\\
6388  &$17\;36\;17.23$&  $-44\;44\;07.8$&  $ 345.56$&  $ -6.74$&  $  9.90$&  $  3.10$&  $  0.37$&  ...\\
6441  &$17\;50\;13.06$&  $-37\;03\;05.2$&  $ 353.53$&  $ -5.01$&  $ 11.60$&  $  3.90$&  $  0.47$&  ...\\
6522  &$18\;03\;34.02$&  $-30\;02\;02.3$&  $   1.02$&  $ -3.93$&  $  7.70$&  $  0.60$&  $  0.48$&  $  0.71$\\
6528  &$18\;04\;49.64$&  $-30\;03\;22.6$&  $   1.14$&  $ -4.17$&  $  7.90$&  $  0.60$&  $  0.54$&  $ -1.00$\\
6544  &$18\;07\;20.58$&  $-24\;59\;50.4$&  $   5.84$&  $ -2.20$&  $  3.00$&  $  5.10$&  $  0.76$&  $  1.00$\\
6553  &$18\;09\;17.60$&  $-25\;54\;31.3$&  $   5.26$&  $ -3.03$&  $  6.00$&  $  2.20$&  $  0.63$&  $ -1.00$\\
6569  &$18\;13\;38.80$&  $-31\;49\;36.8$&  $   0.48$&  $ -6.68$&  $ 10.90$&  $  3.10$&  $  0.53$&  ...\\
6624  &$18\;23\;40.51$&  $-30\;21\;39.7$&  $   2.79$&  $ -7.91$&  $  7.90$&  $  1.20$&  $  0.28$&  $ -1.00$\\
6626  &$18\;24\;32.81$&  $-24\;52\;11.2$&  $   7.80$&  $ -5.58$&  $  5.50$&  $  2.70$&  $  0.40$&  $  0.90$\\
6637  &$18\;31\;23.10$&  $-32\;20\;53.1$&  $   1.72$&  $-10.27$&  $  8.80$&  $  1.70$&  $  0.18$&  $ -1.00$\\
6638  &$18\;30\;56.10$&  $-25\;29\;50.9$&  $   7.90$&  $ -7.15$&  $  9.40$&  $  2.20$&  $  0.41$&  $ -0.30$\\
6652  &$18\;35\;45.63$&  $-32\;59\;26.6$&  $   1.53$&  $-11.38$&  $ 10.00$&  $  2.70$&  $  0.09$&  $ -1.00$\\
6723  &$18\;59\;33.15$&  $-36\;37\;56.1$&  $   0.07$&  $-17.30$&  $  8.70$&  $  2.60$&  $  0.05$&  $ -0.08$\\
6752  &$19\;10\;52.11$&  $-59\;59\;04.4$&  $ 336.49$&  $-25.63$&  $  4.00$&  $  5.20$&  $  0.04$&  $  1.00$\\
7078  &$21\;29\;58.33$&  $+12\;10\;01.2$&  $  65.01$&  $-27.31$&  $ 10.40$&  $ 10.40$&  $  0.10$&  $  0.67$\\

 \hline
  \end{tabular}
  \end{center}
 \end{table*}

\begin{sidewaystable*}

\textbf{TableA2.} The ages and metallicities obtained by three methods (colour,
Lick-index and full spectrum-fitting) and literatures.
The first column lists the names of Galactic GCs. The 2nd to 7th columns are the ages,
metallicities and their errors obtained from three methods, and the 8th and 9th columns
are the same parameters from literatures. In the 10th to 12th columns, we list the 
change of age when eliminating the HB stars for 18 GCs (details see Section 5.1). And the 13th to 15th columns give the change of age when removing the BSs for 9 GCs (details see Section 5.2).
\tabcolsep=0.06in

\label{Tab:frac}
\renewcommand{\arraystretch}{1.4}
\begin{flushleft}
\begin{tabular}{rrrrrrrrrrrrrrrrrrr}
\hline
\hline\noalign{\smallskip}

(1)&(2)&(3)&(4)&(5)&(6)&(7)&(8)&(9)&(10)&(11)&(12)&(13)&(14)&(15)\\
NGC  &$\tau$\tiny${\rm_{C}}$   &[Fe/H]\tiny{$\rm_{C}$} &$\tau$\tiny{$\rm_{I}$}&[Fe/H]\tiny{$\rm_{I}$} 
&$\tau$\tiny{$\rm_{S} $}&[Fe/H]\tiny{$\rm_{S}$} &$\tau$\tiny${\rm_{L}}$&[Fe/H]\tiny{$\rm_{L}$} &$\Delta\tau$\tiny{$\rm^{C}_{HB}$}\normalsize{}&$\Delta\tau$\tiny{$\rm^{I}_{HB}$}\normalsize{}
&$\Delta\tau$\tiny{$\rm^{S}_{HB}$}\normalsize{}
&$\Delta\tau$\tiny{$\rm^{C}_{BSs}$}\normalsize{}&$\Delta\tau$\tiny{$\rm^{I}_{BSs}$}\normalsize{}
&$\Delta\tau$\tiny{$\rm^{S}_{BSs}$}\normalsize{}\\
          &\tiny{(Gyr)} &\tiny{(dex)} &\tiny{(Gyr)}&\tiny{(dex)}&\tiny{(Gyr)}&\tiny{(dex)}&\tiny{(Gyr)}
          &\tiny{(dex)}    &\tiny{(Gyr)} & \tiny{(Gyr)}&\tiny{(Gyr)}
             &\tiny{(Gyr)} &\tiny{(Gyr)}&\tiny{(Gyr)}     \\

\hline\noalign{\smallskip}
 104       &10.00$^{+2.80}_{-2.80}$&$-0.60$$^{+0.28}_{-0.11}$&12.00$^{+3.00}_{-2.00}$&$-0.75$$^{+0.08}_{-0.08}$&7.00$^{+1.20}_{-2.80}$&$-0.60$$^{+0.26}_{-0.09}$&10.70&$-0.76$    & 0.00&$0.00$& 0.00  
  &$0.00$&$+3.00$&0.00\\
 1851       &10.00$^{+5.00}_{-5.00}$&$-1.20$$^{+0.10}_{-0.05}$&15.00$^{+0.00}_{-0.00}$&$-1.70$$^{+0.00}_{-0.00}$&8.00$^{+6.00}_{-3.00}$&$-1.30$$^{+0.28}_{-0.12}$&9.20&$-1.36$    & $0.00$&uncom$^{\ast}$& $+1.00$\\
 1904       &11.00$^{+3.50}_{-2.20}$&$-1.50$$^{+0.18}_{-0.42}$&15.00$^{+0.00}_{-0.00}$&$-2.30$$^{+0.00}_{-0.00}$&10.00$^{+2.40}_{-2.30}$&$-1.70$$^{+0.22}_{-0.10}$&11.70&$-1.69$   & $+1.00$&uncom$^{\ast}$&$+1.00$ 
  &$+2.00$&uncom$^{c}$&$+1.00$\\
 2298       &10.00$^{+2.10}_{-2.00}$&$-1.90$$^{+0.30}_{-0.23}$&15.00$^{+0.00}_{-0.00}$&$-2.30$$^{+0.00}_{-0.00}$&14.00$^{+0.17}_{-2.80}$
 &$-1.90$$^{+0.30}_{-0.30}$&12.60&$-1.97$  \\
 2808       &10.00$^{+2.00}_{-3.20}$&$-1.60$$^{+0.20}_{-0.11}$&10.00$^{+2.00}_{-0.50}$&$-1.45$$^{+0.05}_{-0.05}$&6.00$^{+3.10}_{-0.08}$
 &$-1.40$$^{+0.11}_{-0.18}$&10.20&$-1.37$ & & &  
  &$+2.00$&$+1.00$&$+2.00$\\
 3201       &10.00$^{+3.80}_{-2.30}$&$-1.30$$^{+0.30}_{-0.15}$&15.00$^{+0.00}_{-0.00}$&$-2.30$$^{+0.00}_{-0.00}$&12.00$^{+3.20}_{-2.60}$&$-1.60$$^{+0.07}_{-0.24}$&11.30&$-1.61$   \\
 5286       &9.00$^{+3.80}_{-3.60}$&$-1.80$$^{+0.20}_{-0.31}$&10.00$^{+3.20}_{-2.00}$&$-1.68$$^{+0.03}_{-0.03}$&8.00$^{+4.30}_{-2.10}$&$-1.80$$^{+0.13}_{-0.34}$&12.50&$-1.79$   \\
 5904       &10.00$^{+2.10}_{-2.80}$&$-1.40$$^{+0.10}_{-0.12}$&10.00$^{+2.00}_{-4.50}$&$-1.45$$^{+0.05}_{-0.05}$&9.00$^{+2.20}_{-3.20}$
 &$-1.50$$^{+0.18}_{-0.31}$&10.90&$-1.40$     & $+2.00$&$+5.00$&$+1.00$
  &$+1.00$&$+2.00$&$+2.00$\\
 5927       &10.00$^{+3.50}_{-5.50}$&$-0.60$$^{+0.30}_{-0.05}$&12.00$^{+2.00}_{-2.00}$&$-0.55$$^{+0.10}_{-0.10}$&12.00$^{+1.30}_{-3.20}$&$-0.70$$^{+0.04}_{-0.09}$&10.50&$-0.64$   & $0.00$&$+0.00$&$0.00$
  &$+2.00$&$+3.00$&$+2.00$\\
 5946       &8.00$^{+2.00}_{-2.00}$&$-1.70$$^{+0.08}_{-0.15}$&15.00$^{+0.00}_{-0.00}$&$-1.70$$^{+0.00}_{-0.00}$&14.00$^{+1.30}_{-2.20}$&$-2.00$$^{+0.31}_{-0.26}$&10.10&$-1.54$   & $+1.00$&uncom$^{\ast}$&$0.00$\\
 5986       &8.00$^{+3.70}_{-2.80}$&$-1.70$$^{+0.12}_{-0.42}$&12.00$^{+3.00}_{-1.00}$&$-1.70$$^{+0.00}_{-0.00}$&6.00$^{+7.00}_{-1.60}$&$-1.70$$^{+0.13}_{-0.30}$&12.10&$-1.67$  \\
 6121       &10.00$^{+3.00}_{-4.20}$&$-1.60$$^{+0.18}_{-0.26}$&12.00$^{+1.00}_{-2.00}$&$-1.20$$^{+0.05}_{-0.05}$&13.00$^{+1.90}_{-2.10}$&$-1.50$$^{+0.12}_{-0.40}$&11.70&$-1.33$   \\
 6171       &5.00$^{+10.00}_{-2.00}$&$-0.40$$^{+0.20}_{-0.43}$&12.00$^{+0.80}_{-2.00}$&$-1.10$$^{+0.05}_{-0.05}$&12.00$^{+2.10}_{-3.00}$&$-1.30$$^{+0.11}_{-0.21}$&11.70&$-1.13$   \\
 6218       &9.00$^{+2.00}_{-3.20}$&$-1.70$$^{+0.12}_{-0.42}$&9.00$^{+4.00}_{-3.50}$&$-1.46$$^{+0.03}_{-0.03}$&12.00$^{+2.10}_{-5.10}$&$-1.60$$^{+0.17}_{-0.50}$&12.50&$-1.61$  \\
 6235       &10.00$^{+1.60}_{-3.20}$&$-1.60$$^{+0.18}_{-0.10}$&10.0$^{+2.00}_{-2.70}$&$-1.48$$^{+0.03}_{-0.03}$&12.00$^{+1.00}_{-0.70}$&$-1.40$$^{+0.11}_{-0.15}$&10.20&$-1.38$    & $+1.00$&$+5.00$&$+2.00$\\
 6254       &7.00$^{+1.00}_{-2.00}$&$-1.80$$^{+0.12}_{-0.33}$&15.00$^{+0.00}_{-0.00}$&$-1.70$$^{+0.00}_{-0.00}$&7.00$^{+3.50}_{-1.50}$&$-1.50$$^{+0.10}_{-0.40}$&11.80&$-1.60$  \\
 6266       &10.00$^{+1.60}_{-3.10}$&$-1.60$$^{+0.12}_{-0.20}$&12.00$^{+1.00}_{-2.00}$&$-1.30$$^{+0.06}_{-0.06}$&7.00$^{+1.70}_{-1.20}$&$-1.50$$^{+0.08}_{-0.29}$&10.30&$-1.18$    & $+1.00$&$+3.00$&$+1.00$
   &$0.00$&$+0.50$&$+3.00$\\
 6284       &11.00$^{+2.70}_{-5.20}$&$-1.40$$^{+0.23}_{-0.17}$&12.50$^{+1.50}_{-2.50}$&$-1.25$$^{+0.05}_{-0.05}$&12.00$^{+1.20}_{-2.10}$&$-1.50$$^{+0.13}_{-0.30}$&9.60&$-1.27$    & $0.00$&$+2.00$&$+1.00$\\
 6304       &9.00$^{+5.00}_{-4.00}$&$-0.90$$^{+0.20}_{-0.12}$&12.00$^{+1.50}_{-4.00}$&$-0.65$$^{+0.06}_{-0.06}$&12.00$^{+1.50}_{-1.70}$&$-0.90$$^{+0.12}_{-0.08}$&10.00&$-0.66$    & $+1.00$&$0.00$&$0.00$\\
 6316       &10.00$^{+1.20}_{-2.10}$&$-1.10$$^{+0.20}_{-0.22}$&15.00$^{+0.00}_{-0.00}$&$-0.90$$^{+0.05}_{-0.05}$&12.00$^{+1.20}_{-3.30}$
  &$-1.10$$^{+0.10}_{-0.17}$&...&$-0.90$    \\
 6333       &7.00$^{+4.60}_{-2.30}$&$-2.10$$^{+0.32}_{-0.30}$&15.00$^{+0.00}_{-0.00}$&$-2.30$$^{+0.00}_{-0.00}$&10.00$^{+4.00}_{-3.30}$&$-2.10$$^{+0.23}_{-0.20}$&...&$-1.65$    \\ 
 6342       &5.00$^{+1.30}_{-1.20}$&$-0.50$$^{+0.20}_{-0.10}$&9.00$^{+4.50}_{-2.30}$&$-1.00$$^{+0.05}_{-0.05}$&10.00$^{+2.00}_{-4.40}$&$-1.00$$^{+0.13}_{-0.12}$&10.30&$-1.01$      & $+2.00$&$0.00$&$0.00$\\
 6352       &12.00$^{+4.00}_{-6.00}$&$-0.50$$^{+0.20}_{-0.11}$&8.00$^{+2.00}_{-0.20}$&$-0.73$$^{+0.08}_{-0.08}$&10.00$^{+4.50}_{-4.80}$&$-0.80$$^{+0.23}_{-0.10}$&9.90&$-0.70$     \\

 \hline

 \end{tabular}
  \end{flushleft}
\end{sidewaystable*}

 \begin{sidewaystable*}[h]

 \textbf{Table A2.} Continue
 \tabcolsep=0.06in
 
 \renewcommand{\arraystretch}{1.4}
 
 \begin{tabular}{rrrrrrrrrrrrrrrrrrrr}
 \hline
 \hline

 (1)&(2)&(3)&(4)&(5)&(6)&(7)&(8)&(9)&(10)&(11)&(12)&(13)&(14)&(15)\\
 NGC  &$\tau$\tiny${\rm_{C}}$   &[Fe/H]\tiny{$\rm_{C}$} &$\tau$\tiny{$\rm_{I}$}&[Fe/H]\tiny{$\rm_{I}$} 
 &$\tau$\tiny{$\rm_{S} $}&[Fe/H]\tiny{$\rm_{S}$} &$\tau$\tiny${\rm_{L}}$&[Fe/H]\tiny{$\rm_{L}$} &$\Delta\tau$\tiny{$\rm^{C}_{HBR}$}\normalsize{}&$\Delta\tau$\tiny{$\rm^{I}_{HBR}$}\normalsize{}
 &$\Delta\tau$\tiny{$\rm^{S}_{HBR}$}\normalsize{}
 &$\Delta\tau$\tiny{$\rm^{C}_{BSs}$}\normalsize{}&$\Delta\tau$\tiny{$\rm^{I}_{BSs}$}\normalsize{}
 &$\Delta\tau$\tiny{$\rm^{S}_{BSs}$}\normalsize{}\\
           &\tiny{(Gyr)} &\tiny{(dex)} &\tiny{(Gyr)}&\tiny{(dex)}&\tiny{(Gyr)}&\tiny{(dex)}&\tiny{(Gyr)}
           &\tiny{(dex)}    &\tiny{(Gyr)} & \tiny{(Gyr)}&\tiny{(Gyr)}
 
              &\tiny{(Gyr)} &\tiny{(Gyr)}&\tiny{(Gyr)}     \\
 
 \hline
6356       &9.00$^{+2.60}_{-2.50}$&$-0.60$$^{+0.11}_{-0.13}$&9.00$^{+3.00}_{-1.00}$&$-0.75$$^{+0.05}_{-0.05}$&11.00$^{+0.10}_{-0.10}$&$-0.90$$^{+0.01}_{-0.01}$&...&$-0.74$   \\
   6362       &11.00$^{+2.00}_{-4.10}$&$-1.10$$^{+0.21}_{-0.12}$&10.00$^{+2.00}_{-2.00}$&$-1.15$$^{+0.04}_{-0.04}$&10.00$^{+1.40}_{-4.10}$&$-1.30$$^{+0.12}_{-0.18}$&11.10&$-1.17$      & $0.00$&$0.00$&$+1.00$
  &$0.00$&$+5.00$&$+2.00$\\
  6388       &5.00$^{+5.00}_{-0.60}$&$-0.60$$^{+0.15}_{-0.13}$&10.00$^{+3.00}_{-2.50}$&$-0.75$$^{+0.05}_{-0.05}$&11.00$^{+3.60}_{-5.40}$
  &$-1.00$$^{+0.12}_{-0.14}$&11.50&$-0.68$     & $+1.00$&$+5.00$&$0.00$\\
  6441       &5.00$^{+6.30}_{-0.50}$&$-0.60$$^{+0.15}_{-0.13}$&9.00$^{+4.00}_{-2.00}$&$-0.65$$^{+0.15}_{-0.15}$&10.00$^{+4.80}_{-4.00}$&$-0.90$$^{+0.38}_{-0.12}$&10.80&$-0.65$      & $+1.00$&$+1.00$&$+1.00$\\
  6522       &10.00$^{+3.20}_{-2.70}$&$-1.40$$^{+0.22}_{-0.01}$&9.00$^{+3.20}_{-1.20}$&$-1.40$$^{+0.05}_{-0.05}$&8.00$^{+4.00}_{-2.80}$&$-1.40$$^{+0.20}_{-0.34}$&...&-1.69\\
  6528       &6.00$^{+4.10}_{-1.80}$&$0.00$$^{+0.13}_{-0.16}$&8.00$^{+2.00}_{-2.00}$&$-0.10$$^{+0.12}_{-0.12}$&13.00$^{+0.50}_{-0.50}$&$-0.50$$^{+0.10}_{-0.14}$&...&$-0.10$      \\
  6544       &9.00$^{+0.70}_{-2.80}$&$-1.70$$^{+0.10}_{-0.12}$&15.00$^{+0.00}_{-0.00}$&$-1.50$$^{+0.02}_{-0.02}$&9.00$^{+0.30}_{-3.20}$&$-1.70$$^{+0.04}_{-0.13}$&...&$-1.38$      \\
  6553       &14.00$^{+3.00}_{-2.90}$&0.00$^{+0.00}_{-0.15}$&10.00$^{+2.00}_{-2.00}$&$0.00$$^{+0.20}_{-0.20}$&13.00$^{+2.50}_{-2.60}$&$-0.80$$^{+0.09}_{-0.13}$&...&$-0.20$      \\
  6569       &11.00$^{+1.00}_{-3.20}$&$-1.20$$^{+0.22}_{-0.11}$&10.00$^{+2.00}_{-1.00}$&$-1.06$$^{+0.03}_{-0.03}$&14.00$^{+0.10}_{-1.00}$
      &$-1.40$$^{+0.10}_{-0.05}$&...&$-1.08$\\
  6624       &10.00$^{+4.00}_{-9.00}$&$-0.70$$^{+0.08}_{-0.13}$&10.00$^{+1.00}_{-1.00}$&$-0.73$$^{+0.03}_{-0.03}$&6.00$^{+3.20}_{-3.30}$
  &$-0.60$$^{+0.10}_{-0.13}$&10.60&$-0.70$        & $0.00$&$0.00$&$+2.00$\\
  6626       &10.00$^{+1.00}_{-4.60}$&$-1.70$$^{+0.22}_{-0.13}$&12.00$^{+0.50}_{-6.00}$&$-1.37$$^{+0.05}_{-0.05}$&9.00$^{+4.00}_{-1.80}$&$-1.40$$^{+0.15}_{-0.40}$&12.00&$-1.21$       \\
  6637       &13.00$^{+2.30}_{-2.80}$&$-0.70$$^{+0.12}_{-0.10}$&15.00$^{+0.00}_{-0.00}$&$-0.84$$^{+0.07}_{-0.07}$&8.00$^{+3.10}_{-1.80}$
  &$-0.90$$^{+0.07}_{-0.14}$&10.60&$-0.78$        & $+2.00$&uncom$^{\ast}$&$+2.00$
  &$+1.00$&uncom$^{\ast}$&$+2.00$\\
  6638       &10.00$^{+3.20}_{-4.10}$&$-1.30$$^{+0.22}_{-0.12}$&10.00$^{+2.00}_{-2.00}$&$-1.08$$^{+0.03}_{-0.03}$&12.00$^{+1.80}_{-3.80}$
    &$-1.20$$^{+0.13}_{-0.13}$&...&$-1.08$\\
  6652       &5.00$^{+6.50}_{-0.60}$&$-0.50$$^{+0.20}_{-0.03}$&12.00$^{+2.00}_{-4.50}$&$-1.10$$^{+0.06}_{-0.06}$&10.00$^{+1.40}_{-4.80}$
  &$-1.10$$^{+0.18}_{-0.08}$&11.40&$-1.10 $       & $+1.00$&$0.00$&$0.00$ \\
  6723       &12.00$^{+3.00}_{-3.50}$&$-1.30$$^{+0.30}_{-0.22}$&9.00$^{+3.00}_{-1.00}$&$-1.45$$^{+0.06}_{-0.06}$&11.00$^{+0.90}_{-5.20}$&$-1.40$$^{+0.21}_{-0.13}$&11.60&$-1.10$       \\
  6752       &10.00$^{+4.80}_{-5.00}$&$-1.80$$^{+0.31}_{-0.30}$&8.00$^{+5.00}_{-0.50}$&$-1.68$$^{+0.04}_{-0.04}$&5.00$^{+2.30}_{-1.80}$&$-1.60$$^{+0.02}_{-0.4}$&12.20&$-1.55$       \\
  7078       &8.00$^{+4.20}_{-2.10}$&$-2.10$$^{+0.23}_{-0.20}$&15.00$^{+0.00}_{-0.00}$&$-2.30$$^{+0.00}_{-0.00}$&8.00$^{+0.80}_{-0.50}$&$-2.30$$^{+0.09}_{-0.00}$&11.80&$-2.33$        & $+3.00$&uncom$^{\ast}$&$+4.00$
 &$+1.00$&uncom$^{\ast}$&$+3.00$\\
 \noalign{\smallskip}\hline
 
 \end{tabular}
 \scriptsize
 
 Note.\\
 $^{\ast}$ For GCs beyond Lick indices grids, we do not compare the age after remove the HB stars or BSs and use $'$uncom$'$
to express the change of age.
 \end{sidewaystable*}

\label{lastpage}
\end{document}